\documentclass{aa}  

\usepackage{graphicx}
\usepackage{txfonts}
\usepackage{amsmath}
\usepackage{physics}
\usepackage{xcolor}
\usepackage{float}
\usepackage{hyperref}
\usepackage{threeparttablex}
\usepackage{siunitx}
\usepackage{orcidlink}
\usepackage{bm}
\newcommand*{\Caii}[1]{Ca~\textsc{ii}#1} 
\newcommand*{\Siiv}[1]{Si~\textsc{iv}#1} 
\newcommand*{\Mgii}[1]{Mg~\textsc{ii}#1} 
\newcommand*{\halpha}{H$\alpha$} 
\begin{document}

   \title{Prominence and coronal rain formation by steady versus stochastic heating and how we can relate it to observations}

   \author{V. Jer\v{c}i\'{c}
          \inst{1}\orcidlink{0000-0003-1862-7904},
          J. M. Jenkins
          \inst{1}\orcidlink{0000-0002-8975-812X},
          R. Keppens
          \inst{1}\orcidlink{0000-0003-3544-2733}
          }

   \institute{\inst{1} Centre for mathematical Plasma-Astrophysics, Celestijnenlaan 200B, 3001 Leuven, KU Leuven, Belgium}

   \date{Received ???; accepted ???}

% \abstract{}{}{}{}{} 
% 5 {} token are mandatory
 
  \abstract
   {Prominences and coronal rain are two forms of coronal condensations for which we still lack satisfactory details on the formation pathways and conditions under which the two come to exist. Even more so, it is unclear why prominences and filaments appear in so many different shapes and sizes, with a vertical rather than a horizontal structure or vice-versa. It is also not clear why coronal rain is present in some cases and not in others.}
   {Our aim is to understand the formation process of prominences and coronal rain in more detail by exploring what influence two specific heating prescriptions can have on the resulting formation and evolution, using simulations. We try to determine why we see prominences with such a variety in their properties, particularly by looking at the large-scale topology and dynamics. We attempted to recreate some of these aspects by simulating different types of localised heating. Besides the differences we see on a large scale, we also attempted to determine what the smaller-scale phenomena are, such as reconnection, the influence of resistivity (or lack thereof), the influence of flows and oscillations.}
   {We compared prominences that formed via a steady versus stochastic type of heating. We performed 2.5D simulations using the open-source {\tt MPI-AMRVAC} code. To further extend the work and allow for future direct comparison with observations, we used \textit{Lightweaver} to form spectra of the filament view of our steady case prominence. With that, we analysed a reconnection event that shares certain characteristics with nanojets.}
   {We show how different forms of localised heating that induce thermal instability result in prominences with different properties. The steady form of heating results in prominence with a clear vertical structure stretching across the magnetic field lines. On the other hand, stochastic heating produces many threads that predominantly have a horizontal motion along the field lines. Furthermore, the specific type of heating also influences the small-scale dynamics. In the steady heating case, the prominence is relatively static; however, there is evidence of reconnection happening almost the entire time the prominence is present. In the case of stochastic heating, the threads are highly dynamic, with them also exhibiting a form of transverse oscillation (strongly resembling the decayless type) similar to the vertically polarised oscillations previously found in observations. The fact that the threads in the stochastic heating case are constantly moving along the field lines suppresses any conditions for reconnection. It, therefore, appears that, to first order, the choice of heating prescription defines whether the prominence-internal dynamics are oriented vertically or horizontally. We closely inspected a sample reconnection event and computed the synthetic optically thick radiation using the open-source \textit{Lightweaver} radiative transfer framework. We find the associated dynamics to imprint clear signatures, both in Doppler and emission, on the resulting spectra that should be testable with state-of-the-art instrumentation such as DKIST.}
   {}

   \titlerunning{Prominence and coronal rain formation and how we can relate it to observations}
   \maketitle

%-------------------------------------------------------------------
\section{Introduction}

    Prominences and filaments are plasma condensations found in the solar corona. The two terms represent the same structure but differ depending on whether we observe them over the solar limb or on the disk, respectively. In most cases, prominences have been simply defined as structures two orders of magnitude colder and denser than the corona surrounding them. However, the simple definition hides their true complexity. From observations, we know prominences come in many different shapes and sizes \citep{Berger2014}. In some observations at the limb, the prominences seem to consist of horizontal threads (Hinode views), while in others, the vertical structure and the dynamics due to Rayleigh-Taylor instability in quiescent prominences are emphasised \citep{Berger2014}. A key factor determining the properties of prominences (and filaments) is their magnetic field \citep{Parenti2014}; their overall characteristics significantly differ depending on whether they are rooted in an active region (AR) or in a quiet Sun area. In either case, it is quite probable that they may never reach a truly steady state in which thermal and force balances are achieved, as such they can be highly dynamic structures and exhibit a range of instabilities \citep{Hillier2021, Jenkins2022, Changmai2023}. Another similar and related form of plasma found in the solar corona is coronal rain \citep{Antolin2020}. In reality, the two forms of condensation are very often found together, indicating that their formation and evolution are very likely closely related.

    The question of prominence formation or, more generally, the formation of condensations in the corona has been a long-standing one with multiple potential theories proposed over the last two decades. In a review by \cite{Mackay2010}, three main processes are described, levitation, injection, and evaporation-condensation. Meanwhile, other mechanisms have also been suggested, such as levitation-condensation \citep{Kaneko_Yokoyama2015, Jenkins2021, Jenkins2022} and plasmoid-fed prominence formation \citep{Zhao2022}. Since prominence formation is hard to observe, most of these processes still need to be observationally confirmed and quantitatively compared. In each of the mentioned mechanisms, cold plasma is either brought up directly from the chromosphere (by levitation, injection, or transportation by magnetic islands) or formed within the corona, condensing via the thermal instability (TI) process. In this work, we focus on the latter. 
    
    The trigger of TI is localised heating usually assumed to be localised around the footpoints of the coronal loop supporting the condensed mass \citep{Antiochos1991}. What this localised heating exactly is is not yet known, and just how important this mechanism really is has been shown through numerous research. How the parameters and type of footpoint heating influence the condensation has been extensively studied in 1D hydro models \citep{Antiochos1999, Muller2003, Muller2004, MendozaBriceno2005, Karpen2008, Mikic2013, Johnston2019, Huang2021, Pelouze2022}. \cite{Antolin2008} via a 1.5D coronal loop model studied the differences between a stochastic type of heating \citep[reminiscent of nanoflare type heating,][]{Parker1988} and Alfv\'{e}n wave heating. In their follow-up paper, \cite{Antolin2010} showed that the stochastic type heating leads to the formation of condensations, while Alfv\'{e}n wave heating, due to its uniform heating of the loop, does not. \cite{Kaneko_Yokoyama2017} took a different approach to exploring the influences on TI. They explored the reconnection-condensation model of prominence formation. They studied how the length of the loop influences the critical condition for condensation, namely $L > \lambda_F$, where $L$ represents the field line length and $\lambda_F$ is the so-called Field length \citep{Field1965}. Field length, $\lambda_F$ is defined as a square root of the ratio of thermal conduction and radiative losses. It was first discussed in \cite{Field1965}, where they discussed how thermal conduction can smoothen thermal instabilities, but there is a certain (very small) cut-off wavelength for which it does not develop. \cite{KoyamaInutsuka2004} argued that if this cut-off wavelength is not resolved then such calculations can lead to artificial phenomena, which do not converge with increasing resolution \citep[see also][]{Sharma2010, Hermans2021}. \cite{Kaneko_Yokoyama2017} showed that condensation via TI is triggered after one satisfies the critical condition by lengthening the field line length due to reconnection. The evaporating flows are not the main trigger in their case. By lengthening the field lines, the balancing effects of thermal conduction are less efficient. An analysis done by \cite{Brughmans2022}, furthermore, showed how the background heating (an essential factor for maintaining a hot corona) can have a major influence on the forming condensation. Hence, even the background heating, which is not the primary trigger of condensation (TI), nonetheless influences the formation and morphology of the prominences. 
    
    With this, we see how wide the range of potential mechanisms for triggering TI really is. As there are many mechanisms and factors playing a role, we need more solid constraints. Considering prominences result from such heating mechanisms they are a perfectly observable phenomenon that indirectly encodes those mechanisms. Some of the earlier works used a form of temporally steady heating \cite{Xia2011, Xia2012, Keppens2014} to form the prominence mass. Other works focused on the influence of a more random, pulse-like type of heating, similar to \cite{Antolin2008, Antolin2010} \citep{Karpen2008, Johnston2019, Zhou2020, Li2022, Jercic2023}. %\cite{Jercic2023} performed a 2D simulation showing just how much the formed prominence actually depends on the stochastic formation process, more specifically on the height and the strength of those pulses. 
    Although significant work has been done exploring the formation process more work is still needed. Moreover, comparison with observations is of great relevance to eventually be able to fully understand the formation mechanisms of prominences and coronal rain. 
    
    In observations prominences and filaments appear differently. How exactly we see them, the particular shape and size of the structure depends on the viewing angle but also which particular wavelength we use for the observation. For example, we see prominences in emission and filaments in absorption when observing in Hydrogen \halpha. The filaments observed in H$\alpha$ represent a dark absorption line in comparison to its chromospheric background. On the other hand, the prominences, when using optically cool lines, scatter the absorbed light in all directions and because the background is dark in those wavelengths, we see the scattered light in emission \citep{Heinzel_chapter2015}. Due to the multithermal structure of prominences and filaments, different wavelengths are needed to be able to fully capture all the details. For example, the prominence core can be an optically thick structure when using UV and EUV lines, and it appears dark. However, the relatively (geometrically) thin transition region between the prominence and corona (so-called PCTR), is optically thin to UV and EUV lines, and we can successfully use such lines to analyse the details of the PCTR \citep{Labrosse2010, Parenti2012}.
    
    For the Local Thermodynamic Equilibrium (LTE) to be valid the collisions need to dominate the transfer of energy and the electron population states of a given element. Due to the relatively tenuous nature of prominence plasma, this is not strictly applicable. A more complex approach is needed to solve the statistical equilibrium and subsequent radiative transfer. Although it has received significant attention since the 90s, spectral diagnostics of prominences still have a long way to go. In early attempts, the synthetic spectral lines were created on the basis of (relatively) simple isobaric and/or isothermal prominence slabs \citep{Paletou1993, Paletou1995, Heinzel1995, Gouttebroze2007}. These were followed by 1D and 2D models with more complex pressure and temperature PCTR stratifications \citep{Anzer1999, Heinzel_Anzer2001, Heinzel_etal2001, Heinzel2005, Gunar2007}. With the advent of new space missions such as IRIS, the focus turned more to the specific study of the line formation, what causes it and what properties of the Sun's atmosphere influence it the most \citep{Heinzel2014, Heinzel2015b, Gunar2022}. Regardless of the efforts to create synthetic spectra, they have been only used on simplistic models of solar prominences. With these methods long representing the state-of-the-art, \cite{Heinzel2015a} derived a synthetic visualisation technique for the approximately optically thin Hydrogen H$\alpha$ transition that bypassed the expensive NLTE iterations, the application of which is demonstrated in \cite{Claes2020, Ballester2020, Zhou2020, Jenkins2021, Jercic2023} and in 3D by \cite{Gunar2015}. Recently, \cite{Jenkins2023} used the properties of their simulated atmosphere with a prominence present \citep[based on their earlier work][]{Jenkins2022} to self-consistently synthesise spectral lines commonly employed in the study of solar prominences and filaments (\halpha, \Caii{~H\&K}, \Caii{~8542}, \Mgii{~h\&k}). They applied the \textit{Lightweaver} \footnote{https://goobley.github.io/Lightweaver/} \citep{Osborne2021} framework to create their realistic synthetic spectra, an identical approach to the numerous works on the study of chromospheric spectral lines that analyse their relation to the underlying atmospheric physical properties using similar forward modelling \citep{Carlsson1997, Leenaarts2012, Leenaarts2013a, Leenaarts2013b, Pereira2013, Pereira2015, Bjorgen2018}. \cite{Jenkins2023} used a 1.5D approach to treat every vertical column independently (hence not properly treating scattering), but did allow for the full NLTE dynamics. The advantage of the \textit{Lightweaver} tool is that it can directly interface with a multi-MHD simulation, as we do here. In order to reinforce similar efforts for prominences, we build on the work of \cite{Jenkins2023} and continue to disentangle the connection between the prominence (filament) spectra and the physical conditions within the simulated prominence (filament) atmospheres.

    With this work, we set out to demonstrate the importance of the formation process. We elaborate on the extent to which the heating prescriptions influence the evolution of thermal instability and to what kind of prominence do the coronal condensations later develop. Despite being intricately related, the two forms of condensation, prominences and coronal rain are more often studied separately in numerical simulations. Here, we try to relate the two structures and how their appearance and interaction also depend on the localised heating that initially induces condensation. Moreover, we explore how it can affect even the small-scale details of prominences, such as reconnection and the subsequent formation of plasma blobs and the dynamics of plasma across the field lines. 

    In Section~\ref{ch:methods} we describe the numerical methods used in this work, followed by Section~\ref{ch:results} where we present and describe the results. The following Section~\ref{ch:discussion} gives our discussion in the context of previously described results and in the last section, Section~\ref{ch:conclusions}, we summarise and in a concise manner give the main conclusions of this work.

%--------------------------------------------------------------------
\section{Numerical methods}
\label{ch:methods}

\begin{figure}
        \centering
        \includegraphics[width=\hsize]{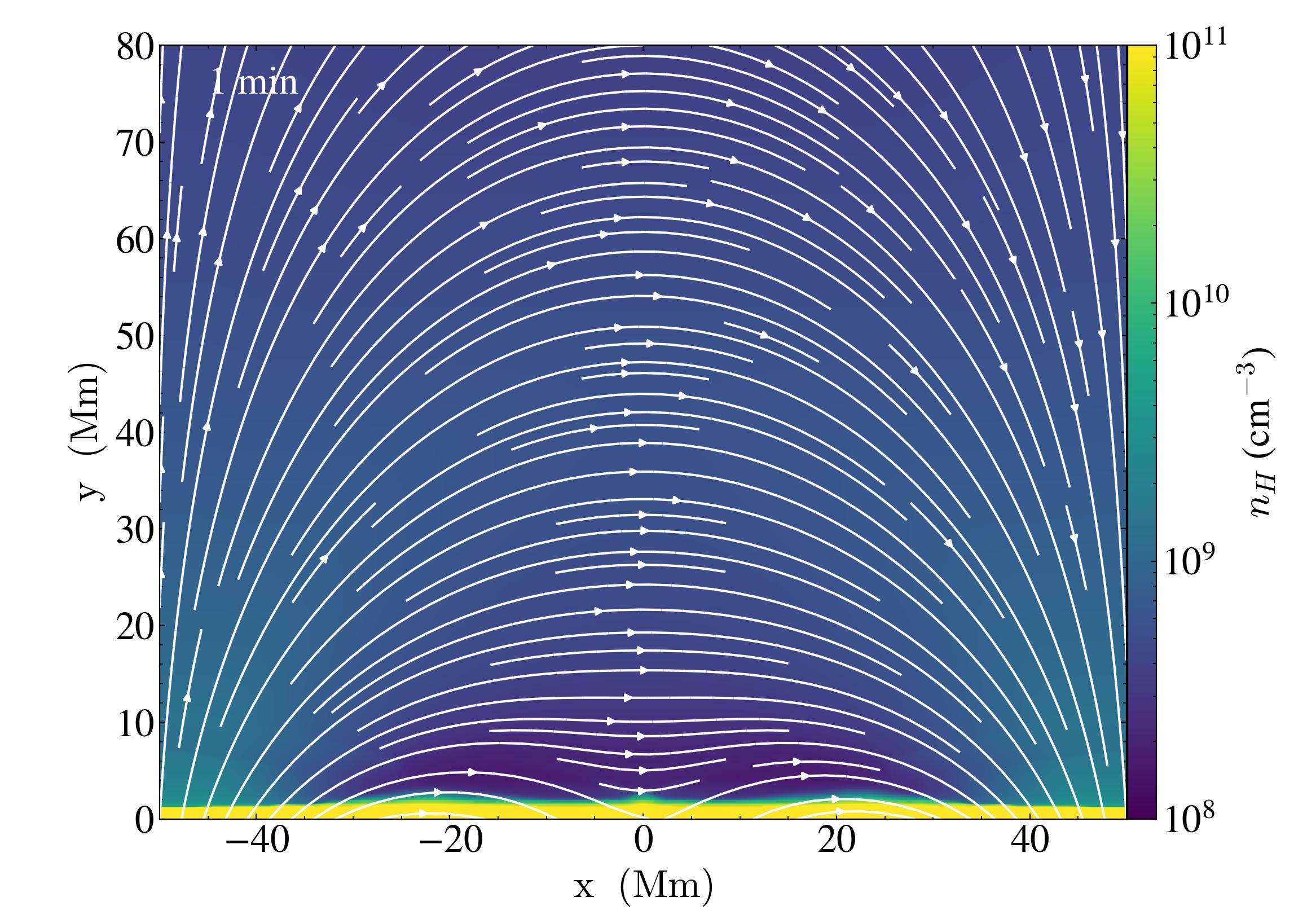}
        \caption{Representations of the field line shape of our quadrupolar magnetic field topology (1\,min into the steady heating scenario).}
        \label{fig:quadrupolar_fieldlines}
    \end{figure}

    We perform a 2.5D simulation including the chromosphere, transition region (TR) and the corona. The domain is a box of 100$\times$80\,Mm with the $x$-axis in the range [-50, 50]\,Mm and the $y$-axis (representing the vertical from the surface of the Sun) [0, 80]\,Mm. The temperature and density stratification are achieved in the same way as in \cite{Li2022} (see their Eq.~8 and 9). For the magnetic field, we adopt a similar quadrupolar topology as was also used in \cite{Terradas2013, Keppens2014, Luna2016}. The 2D representation of the magnetic field lines is shown in Fig.~\ref{fig:quadrupolar_fieldlines}. To improve on the stability during the relaxation phase and to avoid a singularity in plasma $\beta$ ($=2p\mu_0/B^2$), we buried the null point by $y_0=4$\,Mm \citep[same as in][]{Zhang2019}. In the formulae, the field is given by,
    \begin{eqnarray}
        \label{eq:Bx}
        B_x & = & B_0\cos(k_xx)e^{-k_x(y-y_0)} - B_0\cos(3k_xx)e^{-3k_x(y-y_0)} \,,\\
        \label{eq:By}
        B_y & = & -B_0\sin(k_xx)e^{-k_x(y-y_0)} + B_0\sin(3k_xx)e^{-3k_x(y-y_0)} \,,\\
        \label{eq:Bz}
        B_z & = & B_0 \,,
    \end{eqnarray}
    \noindent where $B_0=10$\,G and $k_x=\frac{\pi}{2L_0}$, with $L_0$ being half of the domain width, 50\,Mm. Throughout most of the domain, the magnetic field has a value of $\sim$10\,G, while at the footpoints it reaches $\sim$18\,G.
    
    To perform this study we used an open-source MHD simulation code, MPI\ Adaptive Mesh Refinement Versatile Advection Code ({\tt MPI-AMRVAC}\footnote{http://amrvac.org/}) \citep{Keppens2012, Porth2014, Xia2018, Keppens2021, Keppens2023}. We are solving the following set of non-adiabatic MHD equations,
    \begin{eqnarray} 
        \label{eq:mass}
        \pdv{\rho}{t}+\nabla \cdot (\rho \textbf{v}) & = & 0  \,,\\ 
        \label{eq:momentum}
        \pdv{\rho \textbf{v}}{t} + \nabla \cdot \Bigg(\rho \textbf{vv} + p_{tot}\textbf{I}-\frac{\textbf{BB}}{\mu_0} \Bigg) & = & \rho \textbf{g} \,,\\
        \label{eq:energy}
        \pdv{e}{t} + \nabla \cdot \Bigg(e\textbf{v}-\frac{\textbf{BB}}{\mu_0}\cdot \textbf{v} + \textbf{v}p_{tot} \Bigg) &= & \rho \textbf{g}\cdot \textbf{v} + \nabla \cdot (\pmb{\kappa} \cdot \nabla T)  \, \nonumber \\
        & & - n_Hn_e\Lambda(T) + H \,, \\
        \label{eq:induction}
        \pdv{\textbf{B}}{t} + \nabla \cdot (\textbf{vB} - \textbf{Bv}) & = &0\,.
    \end{eqnarray}
    Here $\rho$ represents density, $\Vec{v}$ velocity and $p_{tot}$ is the total pressure equal to the sum of gas pressure $p = 2.3n_Hk_BT$, and the magnetic pressure $B^2/(2\mu_0)$, $k_B$ is the Boltzmann constant, $T$ is the temperature and $n_H$ and $n_e$ are hydrogen and electron number densities. The plasma is fully ionised with the helium abundance $n_{He}/n_H=0.1$, \textbf{I} is  a unit tensor, $\Vec{B}$ is the magnetic field vector, and $e$ is the total energy density (sum of internal $p/(\gamma-1)$, kinetic $\rho v^2/2$ and magnetic $B^2/(2\mu_0)$ energies). Gravity acceleration is denoted with \Vec{g} and is equal to $-g_{\odot}(R_{\odot}/(R_{\odot}+y))^2\Vec{\hat{y}}$, where $R_{\odot}$ is the solar radius and $g_{\odot}=2.74\times10^4$\,cm\,s$^{-2}$ is gravity acceleration at the solar surface. The coefficient of thermal conduction is $\Vec{\kappa}$ and it only has the component parallel to the magnetic field. The perpendicular component is small enough in comparison that we can ignore it. For $\kappa_{||}$ we take the usual value of the Spitzer conductivity coefficient \citep{Spitzer2006}, 8$\times$10$^{-7}$T$^{5/2}$\,erg\,cm$^{-1}$\,s$^{-1}$\,K$^{-1}$. Optically thin radiative cooling curve is represented by $\Lambda(T)$, which is in our case composed of two parts. For temperatures higher than 10000\,K the values are taken from \cite{Colgan2008} and for temperatures less than 10000\,K the cooling curve is described according to \cite{Dalgarno_McCray1972}. {\tt MPI-AMRVAC} has more than one option of optically thin cooling curves. A more detailed description of each and their comparison can be found in \cite{Hermans2021}. Heating is denoted by \textit{H}, which is in our case composed of background heating, $H_{bg}$ and localised heating. The background heating, $H_{bg}$ is defined as $H_0\exp(-\frac{y}{\lambda_0})$, with $H_0 = 10^{-4}$\,erg\,cm$^{-3}$\,s$^{-1}$, and $\lambda_0 = 50$\,Mm. We include the localised heating after the relaxation phase which is also the moment from which we start counting time, $t=0$. We experiment with two types of localised heating. One is a steady type heating, with the same form as in \cite{Keppens2014} using also the same parameters as they did. The second type of heating we experiment with is the stochastic type heating of a similar form as in Eq.~7 of \cite{Jercic2023}. The difference is that the amplitude used in this study is $E_1 = A(1 + \frac{\tau_i}{\delta\tau_i})$, where $A=0.2$\,erg\,cm$^{-3}$\,s$^{-1}$. As before, $\tau_i$ determines the interpulse duration and $\delta\tau_i$ is the pulse duration. We limit the random position of the pulses to happen only around footpoints, $x_i\in$ [-50, -35] $\cup$ [35, 50]\,Mm and only up to a certain height, $y_i<4$\,Mm \citep[cf. with Eq.~7 in][]{Jercic2023}. The heating length scales for $x$ and $y$, $x_h$ and $y_h$ respectively, are 2\,Mm.
    
    When we later explore the role of resistivity, we instead solve the resistive MHD equations. This is how Ohmic heating in thermal energy enters the total energy density evolution. %I copied the  extra two terms appearing here from the amrvac page: https://amrvac.org/md_doc_equations.html
    \begin{eqnarray}
        \label{eq:res_energy}
        & &\pdv{e}{t} + \nabla \cdot \Bigg(e\textbf{v}-\frac{\textbf{BB}}{\mu_0}\cdot \textbf{v} + \textbf{v}p_{tot} \Bigg) =  \rho \textbf{g}\cdot \textbf{v} + \nabla \cdot (\pmb{\kappa} \cdot \nabla T)  \, \nonumber \\
        & & - n_Hn_e\Lambda(T) + H + \nabla\cdot(\textbf{B}\cross\eta\textbf{J})\,, \\
        \label{eq:res_induction}
        & &\pdv{\textbf{B}}{t} + \nabla \cdot (\textbf{vB} - \textbf{Bv}) = -\nabla\cross(\eta\textbf{J})\,,
    \end{eqnarray}
    \noindent where $\textbf{J} = \nabla\cross\textbf{B}$, is the current and $\eta$ is magnetic resistivity whose value (if present) is uniform across the whole domain. 
    
    In order to solve the equations we use the adaptive mesh refinement (AMR) of {\tt MPI-AMRVAC}. The base level has 200$\times$160 cells and we use 5 levels of refinement, yielding an effective resolution of 31.25$\times$31.25\,km in the smallest cells. The equations (\ref{eq:mass}) - (\ref{eq:res_induction}) are solved using a three-step time discretisation and an SSPRK3 time integrator. For spatial discretisation, a Harten-Lax-van Leer (HLL) approximate Riemann solver \citep{Harten1983} combined with a second-order symmetric TVD limiter \citep{vanLeer1974} was used. For greater stability, we used the magnetic field splitting strategy available in {\tt MPI-AMRVAC} \footnote{\url{https://amrvac.org/md_doc_par.html\#par_mhdlist}}. In other words, the magnetic field is split into a time-invariant part that is handled exactly and a perturbation for which the equations are solved. Furthermore, considering we have a chromosphere and corona with TR we used the transition region adaptive conduction (TRAC) method. More on that numerical method and its influence on the simulation is described in \cite{Jercic2023} (and references therein). In order to control the monopole error (the divergence of the magnetic field approximately zero) we used one of the combined divb cleaning methods that are available in {\tt MPI-AMRVAC} ({\tt lindepowel}). For the left and right boundary at the edges of the $x$-axis, we use symmetric boundary conditions, except for $v_x$ and $B_x$ components, which are asymmetric. The bottom boundary has fixed footpoints, which means the velocity is reflective, and pressure and density are fixed to values calculated according to the hydrostatic equation. The perturbed magnetic field is fixed to zero. At the top boundary, the values of pressure and density are extrapolated according to the gravity stratification, velocity is again reflective and the magnetic field is extrapolated (using a second-order, zero-gradient extrapolation).

    Given that we are including effects of thermal conduction, radiative cooling and localised heating we let the system relax for 90 code time units (approximately 2\,h of real-time). This is required as, even though the equations are in force balance, the discretisation, especially around the TR, is not perfect (which is also why, as mentioned, we use TRAC). However, there are still velocities in our domain even after 2\,h of relaxation (done on a smaller resolution to additionally diffuse any perturbations). These velocities are around 15 km\,s$^{-1}$, and are related to the strong magnetic field we impose (the high value of the $B_z$ component). The largest contribution here is due to the component in the third (ignorable direction, while the velocities in the $x-y$ plane are half the value). Nevertheless, such velocities are all above the TR and the magnitude of the total velocity is smaller than the sound speed.  
%-----------------------------------------------------------------

\section{Results}
\label{ch:results}

    In this work, we experimented with two different types of heating, stochastic versus steady heating in 2.5D, where the two resulted in entirely different types of prominence. We compare the differences and similarities and try to understand how and in what aspects the heating affects the prominence plasma and in what way can our results be related to observations. We start from global characteristics moving on to smaller scales and explaining how even certain small-scale features can be associated with the different types of heating.    

\subsection{Steady heating - Evolution and dynamics}
    \begin{figure*}
        \centering
        \includegraphics[width=\textwidth]{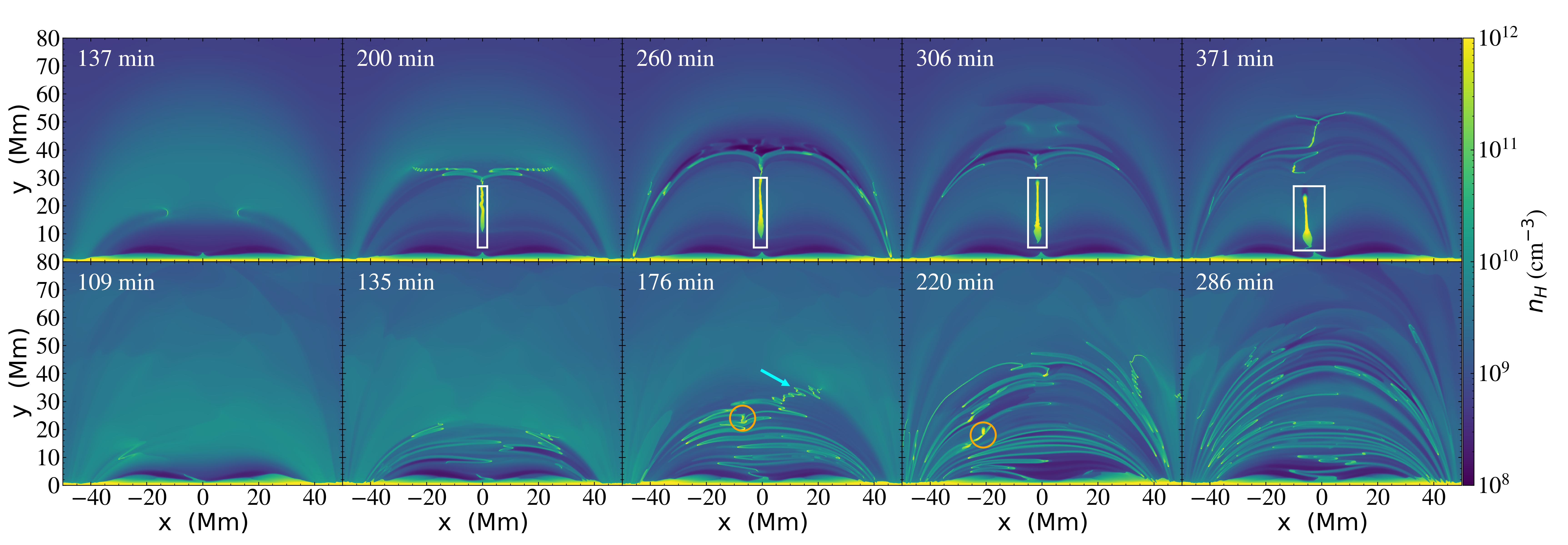}
        \caption{Snapshots in time showing the distribution of $n_H$ in the domain. The top panels display the evolution in the steady heating case, while the bottom panels show the evolution in the stochastic heating case. The white boxes mark the main prominence body in the steady heating case. The orange circles at the two bottom panels mark the same condensation that accumulated a significant amount of mass (for more details see text). The cyan arrow marks an example of fragmentation (for more details see text).}
        \label{fig:steady_stochastic_snapshots}
    \end{figure*}

    \begin{figure*}
        \centering
        \includegraphics[width=\textwidth]{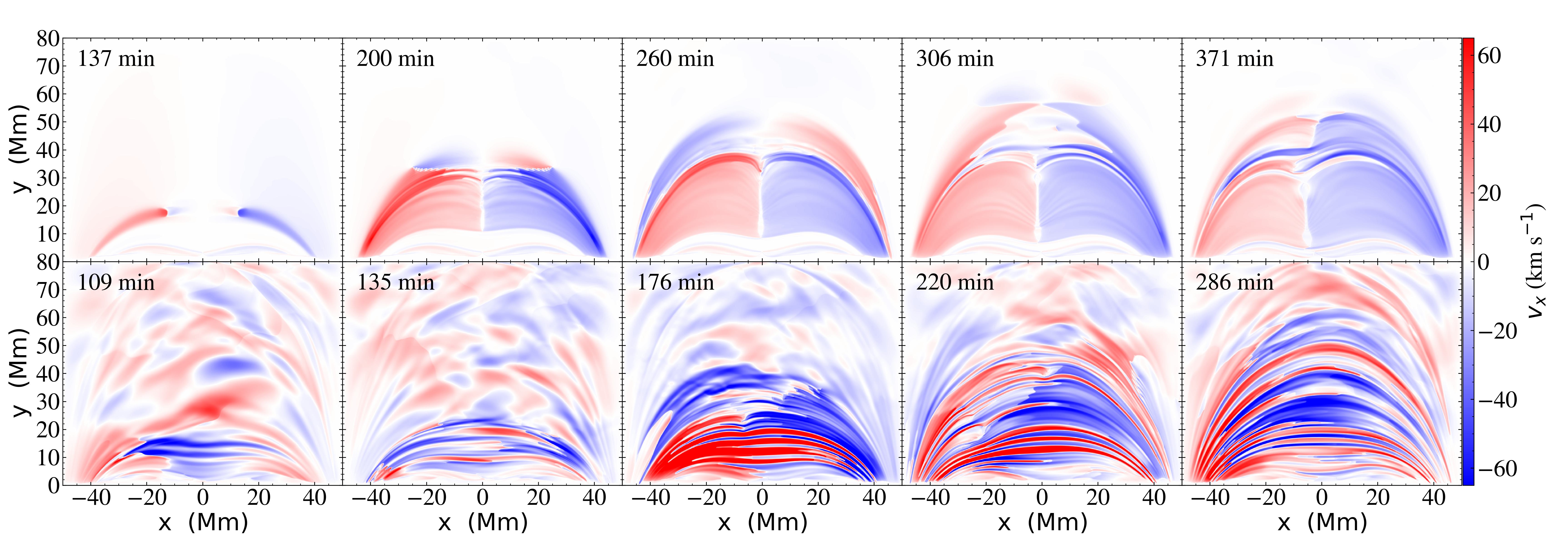}
        \caption{Snapshots in time showing the distribution of the $v_x$ component in the domain. The top panels display the evolution in the steady heating case, while the bottom panels show the evolution in the stochastic heating case.}
        \label{fig:vx_snapshots}
    \end{figure*}
    Approximately 136\,min after we initiate the localised (steady) heating, two condensations at the height of $\sim$18\,Mm form symmetrically around the centre of the domain (see the top leftmost panel of Fig.~\ref{fig:steady_stochastic_snapshots}). Once they are formed there is a region of low pressure around them. However, as there are flows coming from the footpoints that region is asymmetric and ends up being between the two condensing threads as they move closer to each other. The phenomenon of lower pressure close around the forming condensation and the flows resulting from the pressure gradient, have already been described in detail by \cite{Fang2015}. They described siphon flows, driven by the effects of TI and the condensation that forms as a result of it. Approximately 10\,min after, additionally pushed by flows coming from the footpoints, the condensations merge at the middle of the domain, forming a single, steady prominence. As the flows are along the field lines, the $v_x$ component predominantly represents the velocity in the domain. From the top panels of Fig.~\ref{fig:vx_snapshots} we see the evolution of the horizontal flow field. We see that it develops together with the condensation process. When the two condensations finally coalesce in the middle, an outward propagating slow shock is formed \citep[also reported in other works][]{Fang2015, Claes2020, Li2022} on both sides of the now central prominence body. 
    
    As the system attempts to balance the differences in pressure around the forming condensation, more mass is driven into the central part of the condensation and it continuously grows. Furthermore, the flows from the localised heating at the footpoints cause a high ram pressure on the condensation. As a result, the top parts of the condensation that did not yet coalesce start to fragment (second panel of Fig.~\ref{fig:steady_stochastic_snapshots}) due to the thin shell instability \citep[][and references therein]{Claes2020, Hermans2021}. During the evolution, we notice an interaction in the form of reconnection between this top condensation, that is fragmenting and the main prominence body below. As the condensation continuously grows and fragments at its edges, it reaches over the concave down part of the magnetic arcade and drains away, back into the chromosphere (third panel in the top panels of Fig.~\ref{fig:steady_stochastic_snapshots}); the fragmentation eventually results in threads that develop into coronal rain. Around $t=260$\,min the coronal rain touches the chromosphere. As it falls it compresses the plasma ahead of it and leaves a region of lower pressure behind it. The same behaviour of falling condensed plasma has been reported in other numerical works \citep{Fang2015, Li2022}, however the first observational confirmation of it has only recently been reported by \cite[][where they dubbed it the 'fireball' effect]{Antolin2023}. Consequently, once it hits the TR, shocks propagate along the associated field lines and travel up into the corona. It takes about 25\,min for those shocks to reach the top of the loop again. These shocks then trigger another round of condensation that is seen forming around the height of about 48\,Mm (seen on the top two rightmost panels of Fig.~\ref{fig:steady_stochastic_snapshots}). With time, the main prominence body is tipped off balance and starts to increasingly lean towards the left side of the domain, as also seen in MuRAM simulations (Zessner, M. private communication).  
    \begin{figure}
        \centering
        \includegraphics[width=\hsize]{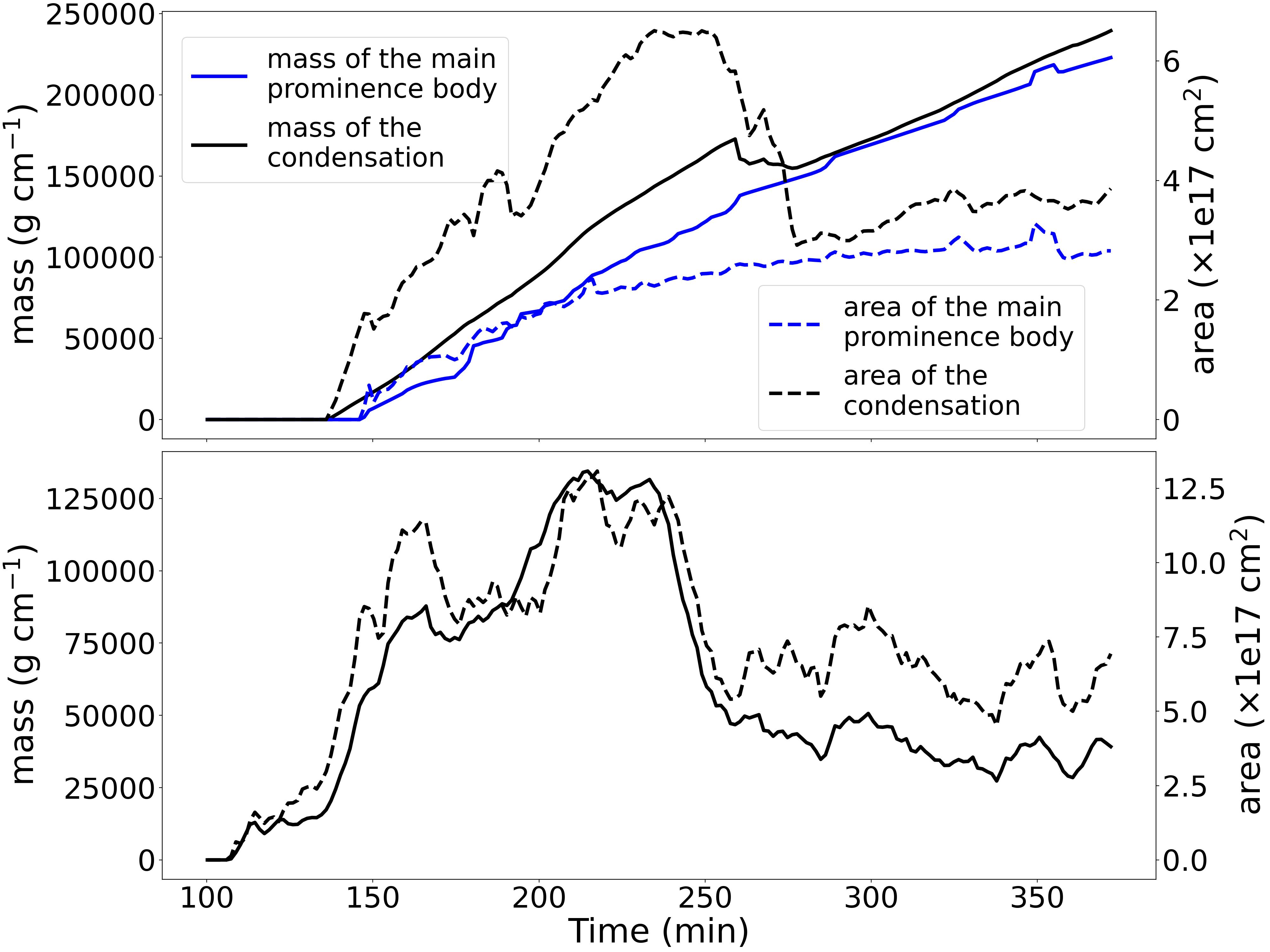}
        \caption{Evolution of the mass and area. The full black line represents the mass of the condensations ($n_H > 10^{10}$\,cm$^{-3}$) in the whole domain but without the TR (everything above 4\,Mm) in the steady and in the stochastic heating case, top and bottom panels respectively. The blue line, for the steady heating case represents the mass of the main prominence body, determined by the same number density threshold and in an area as marked by the white box in Fig.~\ref{fig:steady_stochastic_snapshots}. The dashed lines show the same but for the area of the condensations.}
        \label{fig:mass_area_density_evolution}
    \end{figure}
    
    Figure \ref{fig:mass_area_density_evolution} shows changes in the mass and area of the condensations and this is done for both steady (top) and stochastic (bottom) heating scenarios. The full black line represents (in steady and stochastic case) the condensations in the full domain without the TR (everything below 4\,Mm is excluded). The condensation is defined by a number density threshold of 10$^{10}$\,cm$^{-3}$, as in \cite{Keppens2014}. The dashed lines represent the same but show values of the area. In the steady heating case, we also differentiate the main prominence body (blue) which is defined with the mentioned number density threshold and calculated only for the main body marked as inside the white box drawn in Fig.~\ref{fig:steady_stochastic_snapshots}. We can see that the total mass of the main prominence body continuously grows (full blue line) reaching a value of $\sim2\times10^5$\,g\,cm$^{-1}$ by the end of the simulation. The mass in the coronal domain (full black line) grows until $\sim$260\,min after which the coronal rain drains away. In the next 17\,min a total of 1.8$\times10^4$\,g\,cm$^{-1}$ condensation drains, which equates to approximately 10\% of the condensed mass in the coronal domain before the drainage started (1.7$\times10^5$\,g\,cm$^{-1}$ at $t=259$\,min). After that, most of the mass in the domain is actually part of the main prominence body, even as additional condensations (coronal rain) continuously form into the late stages of the simulation. The area that the condensations in the coronal domain reach (dashed black line) shows a maximum value of more than 6$\times10^{17}$\,cm$^2$. After the main drainage happens, the total area of condensations drops by about 45\% in comparison to the area value before drainage (5.83$\times10^{17}$\,cm$^2$ at $t=259$\,min). The main prominence body appears to reach a plateau in the total area of $\sim$3$\times10^{17}$\,cm$^2$.

\subsection{Stochastic heating - Evolution and dynamics}

    In the case of stochastic heating, condensations start forming at the height of approximately 10\,Mm, about 106\,min after it was initiated. The moment of formation is shown in the bottom leftmost panel of Fig.~\ref{fig:steady_stochastic_snapshots}. The developing condensation continues to grow along the field line in the form of a thread. Similar fragmentation as seen in the steady heating case also happens here. An example is seen on the third panel of the bottom row of Fig.~\ref{fig:steady_stochastic_snapshots} close to $x=20$\,Mm and around the height of $\sim$35\,Mm (marked with the cyan arrow). Here, the condensation had a chance to significantly develop perpendicular to the field lines, simultaneously fragmenting before being pushed along the field lines by the flows travelling along them. The same happens to other existing threads that we see in the simulation. The threads at different locations are pushed in different directions (either up or down along the field lines) by the multitude of stochastic flows coming from the chromosphere and propagating along the field lines. The result is an extremely dynamic system of numerous threads being constantly formed and at the same time, drained away into the chromosphere. In Fig.~\ref{fig:vx_snapshots} throughout the coronal region we see this pattern of field-aligned flows created by the stochastic heating at the footpoints of the magnetic topology. With time, more and more coronal domain is coloured by the pattern of threads moving along the field lines. We see threads occupying ever higher field lines with time from the left to the right panel.  

    The bottom panel of Fig.~\ref{fig:mass_area_density_evolution} describes the changes in mass and area (full and dashed lines respectively) in the stochastic case. Similarly, as in our previous paper \citep{Jercic2023}, we see a more intense initial growth, the mass increases by 1.3$\times10^5$\,g\,cm$^{-1}$ in the first 100\,min of its existence. That equates to a rate of 20.89\,g\,cm$^{-1}$\,s$^{-1}$. The area increased by 1.084$\times10^{18}$\,cm$^2$ in those first 100\,min. As the evolution of condensation progresses certain threads are maintained for a long enough time to accumulate significant amounts of mass. An example is a blob in the bottom fourth panel of Fig.~\ref{fig:steady_stochastic_snapshots} (at x~$\approx$~-20\,Mm and y~$\approx$~20\,Mm, marked with an orange circle) that has been accumulating mass almost since the moment the condensation started. The blob itself was formed at $\sim$169th\,min when two previously existing threads merged. It finally drains, at $\sim$236th\,min which corresponds to the moment we see a significant drop in mass in the bottom panel of Fig.~\ref{fig:mass_area_density_evolution}. The significant drop in mass happens $\sim$150\,min after the first condensation is formed. The mass dropped by 8.5$\times10^4$g\,cm$^{-1}$ in 25\,min, accompanied by an area decrease of 6.185$\times^{17}$\,cm$^2$. After that follows a somewhat steady evolution in both mass and area. 

%-----------------------------------------------------------------
\section{Discussion}
\label{ch:discussion}

\subsection{Steady versus stochastic heating}

    Our steady heating case is built on the example of \cite{Keppens2014}, where the result of such heating is a so-called funnel prominence (large, monolithic plasma structure resting in the dips of the magnetic field configuration). However, even though the localised heating used in our two works is the same, the evolution is not. Unlike \cite{Keppens2014}, we have a slightly different magnetic topology. In our case, the null point is not at $y=0$, but rather at -4\,Mm \citep[similar as in][]{Luna2016, Zhang2019}. We did that in order to ensure greater stability by removing the high plasma-$\beta$ at the null point out of the domain. This also results in somewhat shallower dips as the deeper ones are closer to the TR or buried below (i.e. out of the numerical domain). Furthermore, the magnetic field is stronger in our case, with a value of $B_0=10$\,G rather than 4\,G. Our background heating is weaker \citep[in][they used $H_0 = 3\times10^{-4}$\,erg\,cm$^{-3}$\,s$^{-1}$]{Keppens2014} and the initial TR is at 2\,Mm rather than 2.7\,Mm as in \cite{Keppens2014}. Both affect the temperature distribution of the atmosphere making the shoulders at a lower temperature than the magnetic dip. Regions of even only slightly lower temperatures trigger radiative cooling to act more efficiently there, eventually resulting in the catastrophic cooling effect. On top of all that, there are additional numerical differences between our work and \cite{Keppens2014}, such as a different cooling table, use of the TRAC method \citep{Johnston2020, Zhou2021}, as well as the usage of the so-called magnetic field splitting strategy \citep{Xia2018}, that all contribute to the differences in the general evolution of the funnel prominences in our two works. 
    
    The mass in the steady heating case continuously accumulates onto the main prominence body. Here, the main prominence body reaches a mass of approximately 2$\times$10$^5$\,g\,cm$^{-1}$ which is the maximum value also reached in \cite{Keppens2014}. The difference is that in our case, the main prominence body reached that in a bit less than 4\,hr, while in \cite{Keppens2014} it reached the same value in a bit less than 8\,hr. The prominence in our case has almost double the growth rate than in \cite{Keppens2014}. Furthermore, the maximum area the main prominence body in our case reached, is $\sim$3$\times$10$^{17}$\,cm$^2$, while in \cite{Keppens2014} it is $\sim$10$\times$10$^{17}$\,cm$^2$. We anticipate that the main reasons for these discrepancies are the aforementioned differences in our numerical implementations. TRAC and the different resolutions used in our two works have an influence on the accumulated mass and area. In particular, TRAC regulates the exchange of energy between the chromosphere and corona, hence also the evaporated mass; the method explicitly aims to increase the amount of mass liberated from the chromosphere during evaporation. Since radiative losses are proportional to the square of the number density, this also largely explains the difference in timeframes. Secondary to this, the influence of coarser resolution has been well studied by \cite{Hermans2021}. Their Fig.~8 and 9 show how much the resolution influences the density, surface mass and surface area. As for the plateau value in the area growth seen in the top panel of Fig.~\ref{fig:mass_area_density_evolution}, that can be explained by the influence of the flows strongly compressing the main prominence body and not allowing its further expansion, compounding the aforementioned influence of the resolution.

    The stochastic heating case was inspired by our previous work \citep{Jercic2023}. Here we use the same type of heating (see Section~\ref{ch:methods}), with the change in our set-up orientation enabling the threads to develop in the vertical direction, allowing us to see this type of evolution from a different perspective. Moreover, that difference also allows for a different evolution of mass and area. In the 2D simulation of our previous work, there was a steady accumulation of mass and area of condensation. The vertically rigid magnetic field in our previous work, as well as the deep dip both allowed for a long and steady accumulation of mass and growth in the area. In that sense, the stochastic heating in our previous work resembled more the steady heating in this work. In \cite{Jercic2023} the reason for steady accumulation was the type of the domain, that is the fixed magnetic field topology, rather than the type of heating, as it is in this case. 

    \cite{Pelouze2022} analysed a range of 1D simulations changing parameters of the localised heating and of the loop geometry (resulting in 9000 different simulations). Their goal was to study the mentioned TNE and try to grasp why some TNE cycles produce an abundance of coronal rain while others very little or not at all. They showed that whether a prominence or a TNE cycle (with or without coronal rain) appears depends on the symmetry of both the loop and the heating. Symmetric heating and symmetric geometry build very heavy, massive prominences that heavily weigh on the field lines (though this weighing effect is not incorporated in 1D fixed field hydro models). On the other hand, matching and explaining the stochastic type of heating with any from the range of simulations \cite{Pelouze2022} did, is more complex. Our localised heating is stochastic and impulsive which are characteristics of the heating not taken into account by \cite{Pelouze2022}. If we only consider the fact that we have asymmetric heating and a symmetric loop geometry, then the multitude of condensation and drainage we get (bottom panels of Fig.~\ref{fig:steady_stochastic_snapshots}) is in accordance with the results of \cite{Pelouze2022}. Coronal rain appears in symmetric loops for a broader range of heating variations, as the chances are higher that the condensations have time to form before they drain as a result of any asymmetries in the heating. What we present here thus demonstrates a region of parameter space not fully covered by \cite{Pelouze2022}, and motivates a future extension to their already extensive statistical study. The higher dimensionality of our study, and the adopted quadrupolar field topology, are aspects that are intrinsically enriching the outcome.
    
\subsection{Forming condensations and their resulting mass and area}
    
\begin{figure}
        \centering
        \includegraphics[width=\hsize]{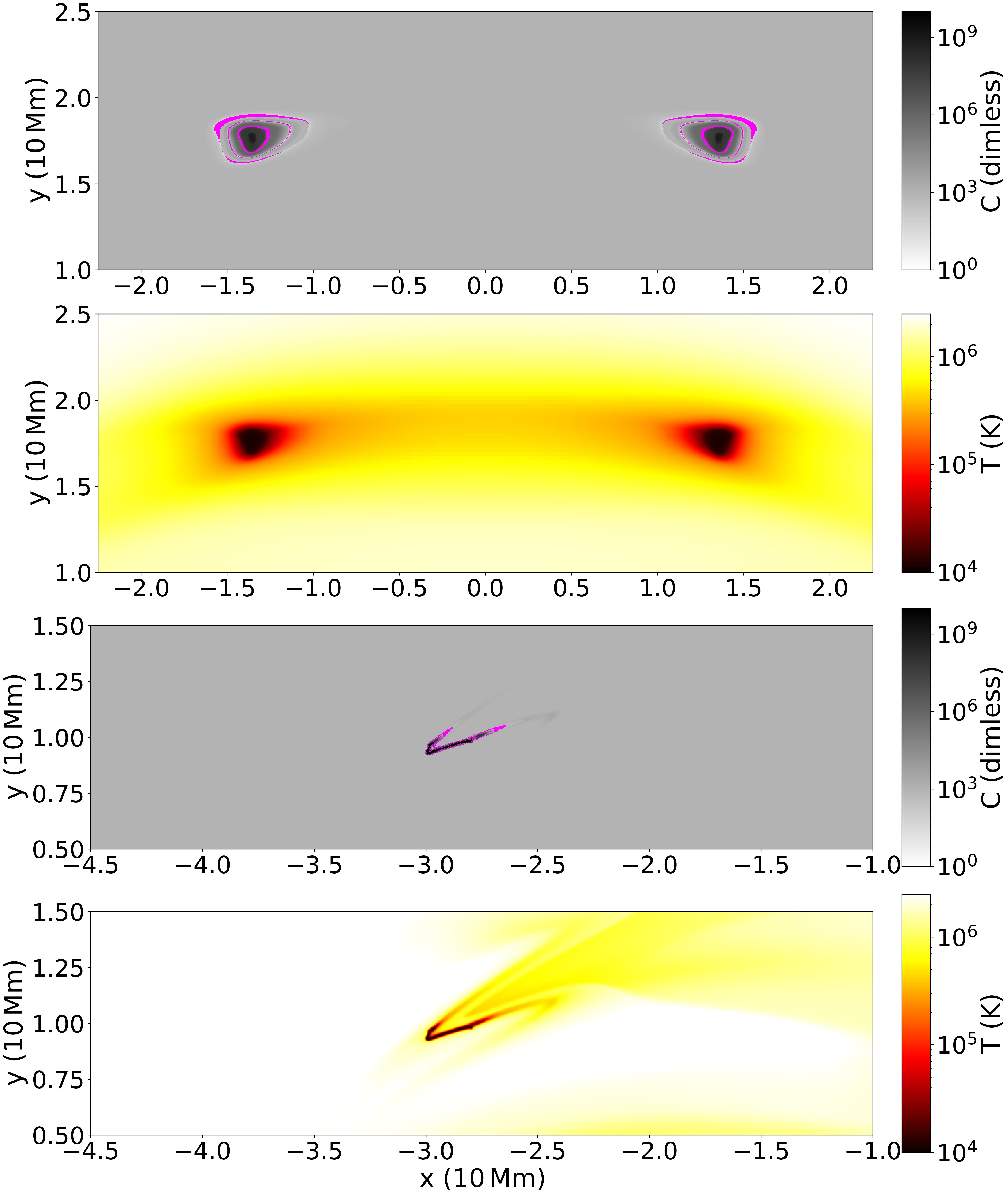}
        \caption{Snapshots showing the dimensionless value of the isochoric criteria for TI and showing the temperature for the same time moment. The first two panels show time $t=136$\,min for the steady heating and the bottom two panels show time $t=109$\,min for the stochastic heating case. Magenta coloured parts on the first and third panels show areas susceptible to TI according to the isochoric assumption (negative values).}
        \label{fig:isochoric_condi}
    \end{figure}

    To further explore the formation of condensation we calculated and plotted the conditions for TI under the isochoric assumption derived by \cite{Parker1953}. \cite{Parker1953} derived the condition by only analysing the energy equation. \cite{Field1965} pointed out that in such a case there is an imbalance in the force equation, and a strong pressure deficit forms because the density does not change. So \cite{Field1965} derived the isobaric condition for the TI analysing only the force equation. Later works also looked into these conditions. \cite{Xia2011} noted that in the initial stages of TI the temperature and pressure change rapidly while the density does not change significantly. In that respect, the isochoric assumption better approximates the initial evolution of TI. \cite{Moschou2015} and \cite{Jenkins2021} again showed the same, with a more complex picture painted by \cite{Brughmans2022} but nevertheless demonstrating isochoric behaviour in the early stages of condensation development. In our steady and stochastic heating, TI is the key factor in the formation of prominence. In Fig.~\ref{fig:isochoric_condi} we plot the isochoric condition for TI for the steady and stochastic heating (first two and last two panels, respectively) with the magenta areas marking the negative values of the isochoric condition (thermally unstable regions). The regions that the isochoric assumption marks as regions of condensation-to-be significantly differ in the steady and stochastic heating case. In the steady case, they describe concentric ellipse-like regions, while in the stochastic heating case, they are tightly aligned with the field line. In time, that can then be related to the evolution of the condensations that follow. In the steady heating case we see the prominence developing across the magnetic field lines while in the stochastic heating case, all the condensations are aligned with the field lines. Hence, we see the type of heating plays an important role in determining how and where thermally unstable regions appear, consequently, it influences how the condensation develops in terms of shape and area.   

    We already pointed out that the growth rate of the prominence mass in the stochastic heating case equates to 20.89\,g\,cm$^{-1}$\,s$^{-1}$. The growth rate of the main prominence body in the steady heating case, considering the first $\approx$120\,min (from its formation to just before coronal rain falls back to the chromosphere) is 23.03\,g\,cm$^{-1}$\,s$^{-1}$. If we assume in both of those cases, that the prominence has a thickness of at least 1\,Mm that adds up to a growth rate of $\approx$2$\times$10$^9$\,g\,s$^{-1}$. If we compare that to Fig.~5 of \cite{Liu2012} and the changes in mass after the prominence reforms, starting at around 08:00 up until 14:00 the growth rate equates to $\approx$2.4$\times$10$^9$\,g\,s$^{-1}$. Another example of a study where they track the mass of a forming prominence is \cite{Berger2012}. From their Fig.~5, we can again calculate the growth rate of the condensation starting at around 22:00 up to 04:00 the next day. The growth rate equates to 3.7$\times$10$^9$\,g\,s$^{-1}$. We see that in both cases the growth rate here matches the order of the growth rates from observations, particularly well with the prominence analysed by \cite{Liu2012}. 

    Comparing the top and bottom panels of Fig.~\ref{fig:mass_area_density_evolution} we notice the mass and area change differently in the stochastic heating compared to the steady heating case. The significant flows throughout the domain are responsible for pushing the material around, not allowing the material to remain within the dip. Condensations in the stochastic case are more free to drain, hence allowing for shorter time scale variations in the mass and area during the entire evolution. The main mass and area increase (see around 230\,min on the bottom panel of Fig.~\ref{fig:mass_area_density_evolution}) happens as certain blobs are longer lived, and consequently manage to accumulate a significant amount of mass. After they drain (after a period where they were in overall balance due to flows maintaining them there), around 250\,min the mass noticeably drops. As the heating supports a constant renewal of condensations, we expect a similar increase in the average mass to repeat, which would then represent a thermal-nonequilibrium (TNE) cycle \citep{Klimchuk2019, Antolin2020, Antolin&Froment2022}. 
    
    It is quite obvious that the topologies resulting from the two initial condensations shown in Fig.~\ref{fig:isochoric_condi} are completely different. The prominence resulting from the steady heating is predominantly characterised by its main, vertically elongated body, with coronal rain coming down from its top following the field lines. A topologically similar prominence was reported in observations by \cite{Antolin2021} and the same one again by \cite{Kumar2023} where it was thoroughly studied in the context of nanojets and a failed eruption event. Furthermore, \cite{Chen2022} also reported on a similar prominence as in our steady case, with the difference that one footpoint of the magnetic topology was rooted in a sunspot region making the surrounding magnetic field stronger than what we simulate. As for the stochastic heating and the threaded prominence formed in such a way, an obvious topological resemblance is found with observations as reported by \cite{Okamoto2016}. There the horizontal flows of the threads are the predominant observable motion and it highly resembles our stochastic heating simulation. Our two different localised heating prescriptions with significantly different types of prominences resulting from it can be clearly related to the prominences found in solar observations. 

\subsection{Transverse oscillations}
\begin{figure}
        \centering
        \includegraphics[width=\hsize]{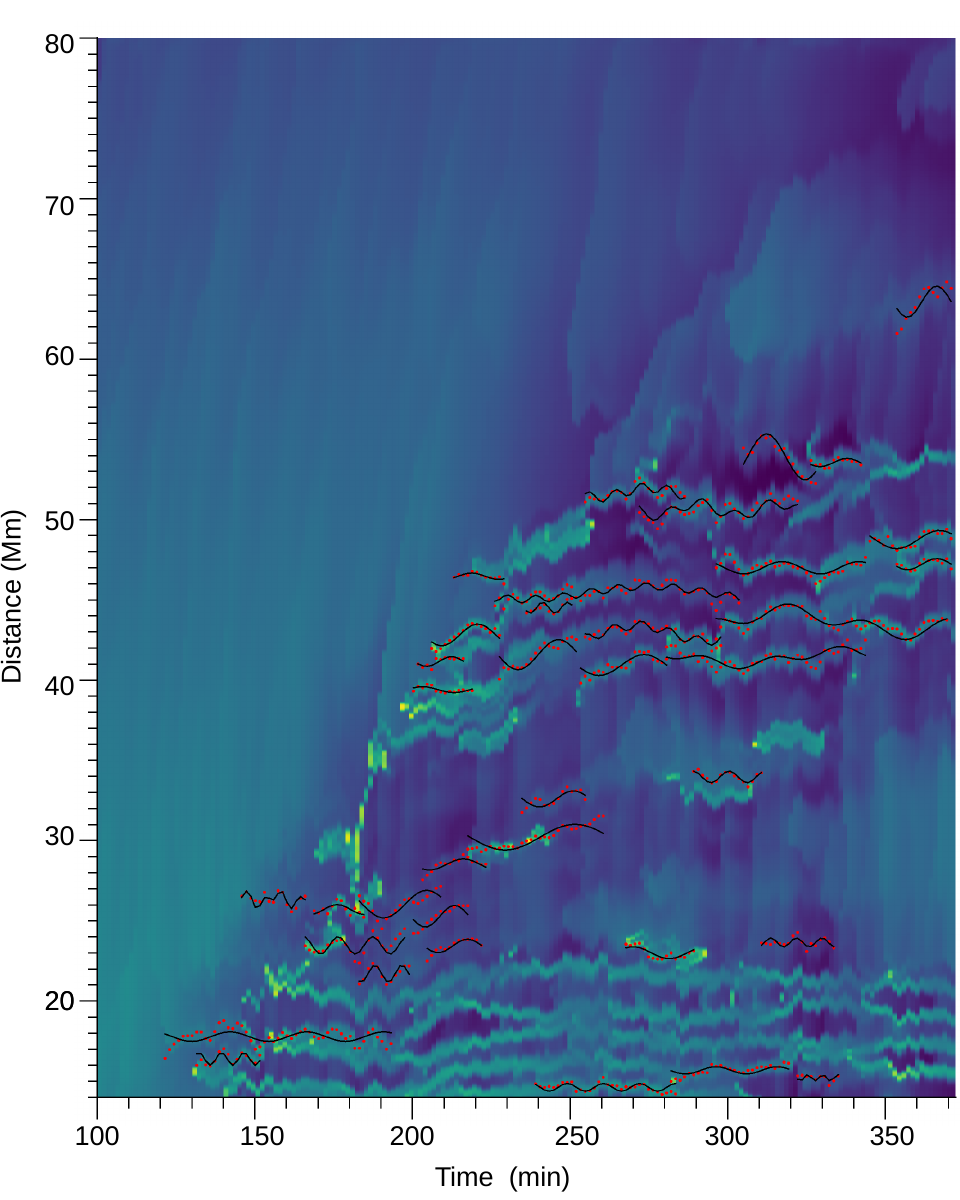} %time_number_density_xcut_0Mm.jpg
        \caption{Time-distance diagram of a cut taken at $x=0$, where the $x$-axis starts at the moment just before the condensations form (at t$\approx$100\,min, until the end of the simulation). Marked with red dots are threads traced by NUWT. Full black line represents a wave fitted by NUWT.}
        \label{fig:decayless_oscill}
    \end{figure}

\begin{figure}
        \centering
        \includegraphics[width=0.9\hsize]{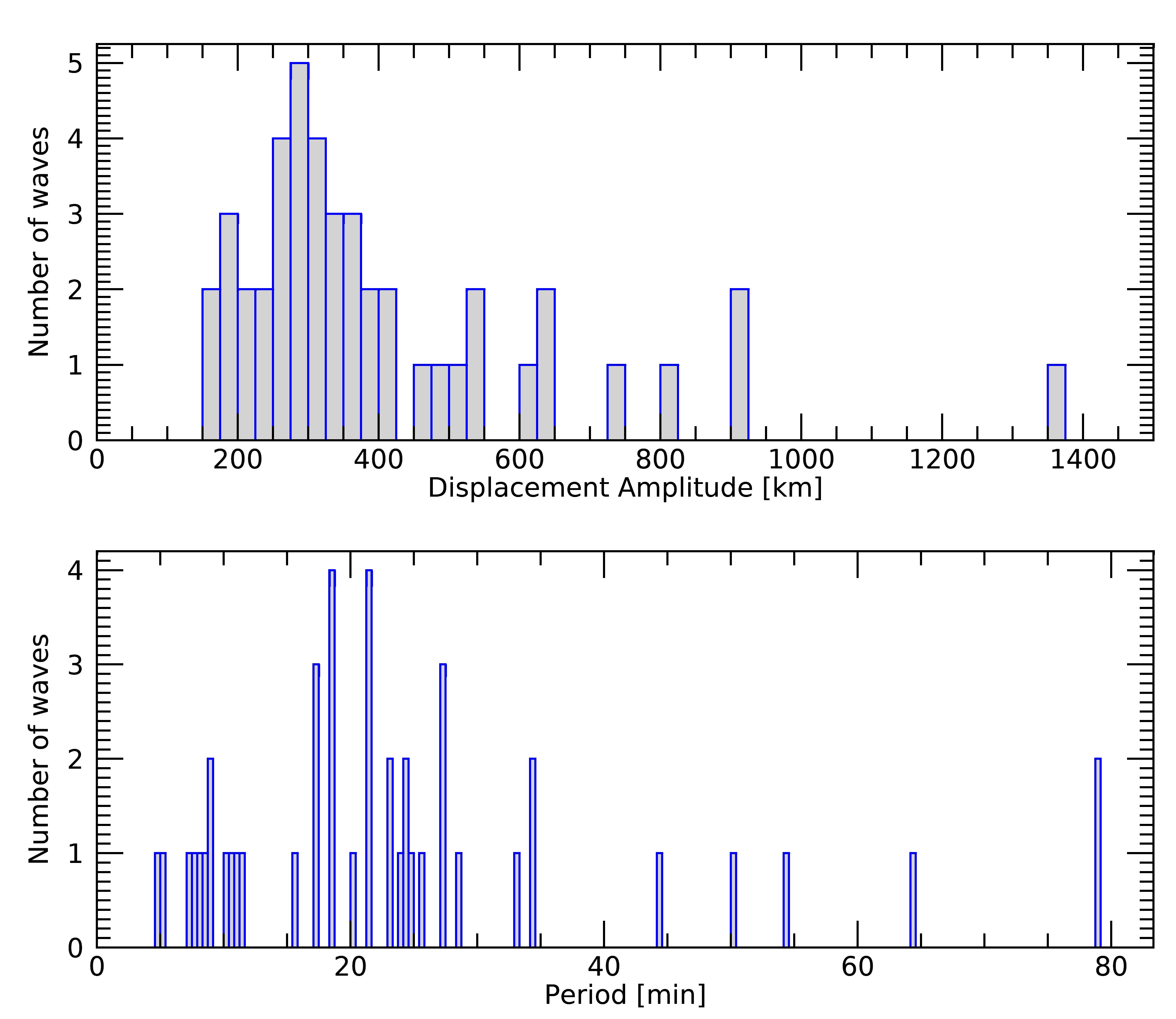} 
        %Histogram from NUWT
        \caption{Histogram of the amplitudes and periods of NUWT fitted oscillations.}
        \label{fig:histogram}
    \end{figure}

    In the bottom panels of Fig.~\ref{fig:vx_snapshots} we presented values of the $x$-component of the velocity for the stochastic heating case. However, even though it represents the dominant motion it is not the only type of motion we see in the domain. To explore how much (or even if) the threads show any kind of motion in the vertical direction (which is nearly perpendicular to the field lines), we plotted a time-distance diagram of a cut through the domain taken at $x=0$. The result is shown in Fig.~\ref{fig:decayless_oscill} and what the threads appear to exhibit are transverse oscillations. 
    The motion of threads we see in Fig.~\ref{fig:decayless_oscill} has been tracked and analysed using the Northumbria Wave Tracking Code (NUWT) \footnote{\url{https://github.com/Richardjmorton/auto_nuwt_public}} \citep{Morton2013, Weberg2018}. The code searches for local maxima, where the end user is responsible for specifying certain parameters (such as the gradient cut-off value for selecting local maxima in the time-distance plot). The particular spatial resolution and time cadence of the data are also fed to NUWT to optimally find the local maxima at every moment. After that, a Gaussian model is fitted to the maxima and the peak locations are tracked in the time-distance diagram that gives the located threads as presented in Fig.~\ref{fig:decayless_oscill} (red dots). We are considering threads located above $y=14$\,Mm. We do not consider the TR nor any low lying threads that are still directly connected to the chromosphere as our goal here is to focus on the motion of the threads. After tracking the threads, NUWT can automatically do a Fourier analysis. In our case, NUWT tracked in total 58 threads with 62 waves. Some threads show oscillations that are a composite of a primary and a secondary wave (in total 8 such threads in our analysis). By overplotting the fitted wave over the extracted data points, we noticed that the fitting did not work well for all the threads. To avoid erroneously fitted oscillations we removed 17 out of the 62 waves. Figure ~\ref{fig:decayless_oscill} shows the 45 waves that we considered well fitted by NUWT's FFT analysis. The histogram of the amplitudes and the periods  of the plotted oscillations is presented in Fig.~\ref{fig:histogram}. We see that the distribution is spread out but that most waves have amplitudes of about 300\,km, and periods of about 20\,min. For the 45 waves we analysed, the mean value of the period is 22\,min, while the median is 19\,min. Transverse oscillations in the solar corona are a highly important topic especially of those oscillations that exhibit a decayless character. A further, more thorough analysis of these oscillations is still needed, however our goal here is to give an overview of the oscillations, and a more detailed analysis is out of the scope of the paper.
    
    \cite{Anfinogentov2015} did a statistical study, and concluded that decayless oscillations are ubiquitous in the solar corona, and they are not necessarily related to an external driver. The average length of the loops they studied was ($219 \pm 117$)\,Mm. The average amplitude and period measured were ($170\pm 100$)\,km and ($4.2\pm2.2$)\,min. They explain these oscillations as standing natural oscillations of the studied coronal loops. However, this is not the only possible source of such oscillations. \cite{Verwichte2017} studied downfalling coronal rain with IRIS in \Siiv~and with Hinode/SOT in \Caii{~h}. They noticed that after the formation, the loop relatively quickly fills up to 60\% of its length with plasma. The amplitudes of the vertically polarised oscillations they measure are ($220\pm 70$)\,km with a period of ($2.6\pm 0.1$)\,min and ($150\pm 70$)\,km with a period of (2.1 $\pm$ 0.1)\,min. They concluded that what they observed was a link between downflows and the period of the loop’s oscillations (as a decrease in mass results in a decrease of the period), however, not a link between the coronal rain and the oscillation phase. Additionally, prior to this study, \cite{Kohutova2016} did an observational study of transverse oscillations of coronal rain, and they found two oscillation regimes. The first is short-period oscillations that they interpret as excited by a standing wave. The measured period of 3.4\,min matches the period of a fundamental harmonic of a loop of length 129\,Mm, estimating the Alfv\'{e}n speed to be around 1000\,km\,s$^{-1}$. The second oscillation regime they find, more relatable to what we observe in our simulation, are large-scale oscillations with amplitudes of around 1\,Mm and a period of 17.4\,min. Due to the long period they measure they interpret such oscillations as a result of a propagating wave. The source of the wave they suspect to be found in a yet undefined mechanism localised at the footpoints of the coronal loop. 
    
    The threads in our simulation exhibit transverse oscillations that display decayless characteristics, as seen in Fig.~\ref{fig:decayless_oscill}. If we consider the fundamental mode of the loop with Alfv\'{e}n speed close to 900\,km\,s$^{-1}$ and the sound speed going up to 300\,km\,s$^{-1}$ (as determined from two randomly chosen points in the loop at $t=176$\,min and $t=220$\,min in the bottom panels of Fig.~\ref{fig:steady_stochastic_snapshots}) while approximating the loop as a half-circle of radius 40\,Mm the periods are $\sim$2\,min and 7\,min respectively. This range of values is more closely related to the first regime of oscillations as found in \cite{Kohutova2016}. However, this is not close to the values we measure. On the other hand, if we consider the values of Alfv\'{e}n and sound speed as measured in the condensation (for example the blob as seen in the orange circle in Fig.~\ref{fig:steady_stochastic_snapshots} at $t=176$\,min and $t=220$\,min in the bottom panels) the values are closer to the range of tens of km\,s$^{-1}$ in which case the fundamental standing mode would have periods closer to what we measure with NUWT (in the range of tens of minutes). However, as the entire loop is not actually filled with condensations, what we believe to be a plausible explanation for these vertical oscillations, is the influence of the threads as they condense and accumulate mass. Furthermore, to fully complete the analysis we compare our simulation to an analytical model, the dispersion relation as described by Eq.\,27 in \cite{DR2005}. The authors there are considering a thin prominence fibril inside a loop of a density different from the one of the corona. Their calculations imply a highly idealised setup ($\beta=0$, no gravity and no curvature). Using Eq.\,27 from \cite{DR2005} we got a value of the period of 13\,min for a thread of half length of 5\,Mm, the ratio of density inside the loop and coronal density 3. The ratio of the thread and the coronal density 220. The length of the total loop we approximated as half circle with a radius of 40\,Mm. Lastly, we took Alfv\'{e}n velocity to be 1490\,km\,s$^{-1}$ (a randomly chosen point in the coronal part of the domain). We note that the 13\,min prediction for the period is in the lower end of our actually identified period distribution, another indication that pressure variations (ignored in the analytic model) are paramount. 
    %We could also apply the dispersion relation as the Eq.\,27 in \cite{DR2005} where they consider a thin prominence fibril inside a loop of a density different from the one of the corona. However, their calculations imply a highly idealized setup ($\beta=0$, no gravity and no curvature). Considering the length of our simulation and the multitude of threads appearing in the simulation it is difficult to estimate the range of values we could get with Eq.\,27 of \cite{DR2005}. The mentioned equation, besides on the length of a thread, depends also on the density in the loop and the density of the corona. Already the length of the thread is very diverse through the simulation; going from very small threads of 2-3\,Mm length up to threads that are extending along most of the loop length. Similar is with the ratio of the density in the loop and in the corona, as well as the ratio of the prominence density and coronal density. \cite{Terradas2008} applied the same dispersion relation to observations of \textit{Hinode}, however, even they warn on the complexity of real prominences. Similarly, in our simulation we have an inhomogeneous distribution of the Alfv\'{e}n velocity, due to the variations of the magnetic field and density. As a result, the range of values we could get with the Eq.\,27 from \cite{DR2005} is large and does not offer us a straightforward conclusion on the oscillations.
    
    As the condensations (the threads) develop, they inevitably dip the field lines. The Lorentz force then strengthens, trying to restore the original shape of the field lines and oscillations are initiated. The dynamics of threads are additionally influenced by the flows coming from the footpoints affecting the velocities we measure in the domain. Even though \cite{Verwichte2017} were not yet able to make a clear connection to the source of the observed vertically polarised oscillations we are now able to go one step further. Considering also the amplitudes and periods we measure and the long period oscillations that were measured by \cite{Kohutova2016} we can connect the transverse oscillations in our simulation as directly caused by the coronal rain dynamics and further, indirectly caused by the stochastic localised heating at the footpoints \citep[considering the localised heating influences the dynamics of the coronal rain as was shown in][]{Jercic2023}. An additional matter to mention is that \cite{Verwichte2017} and \cite{Kohutova2016} studied observations while we are analysing density variations. A closer comparison might be achieved by doing synthetic observational images of our data. However, since the common synthesis method is almost directly related to the density, we do not expect our current results to significantly change.
    
    We support here the conclusion that the decayless nature of the transverse oscillations of individual (rain) blobs on the arcade is here mostly related to the continuous stochastic heating that is enforced, and the possible constructive and destructive interference of waves traversing the multi-structured arcade. However, we still need to point out certain disadvantages of 2.5D simulations and the analysis of such. This type of 2.5D nonlinear simulation is challenged to resolve the details in the cross-field response for example due to the resonant absorption \citep{Arregui2011}. As a result, we cannot fully disregard the possibility that our oscillations are decayless as there are no damping mechanisms, more specifically we cannot say much on the influence of resonant absorption on our threads. We need to consider the fact that the decayless characteristic we see is influenced by the lack of damping mechanisms. In a 3D simulation this aspect could be additionally explored, and more future work on this topic is needed. It would be interesting to explore if we would still see the same decayless character of the threads in a 3D simulation with the same type of localised heating. We leave the answer to this question for future work.
    % The reason why Anfinogentov+2015 interpreted their oscillations as a standing one is that they found a correlation between the oscillation period and the length of the oscillating loop.
    
\subsection{Reconnection}
\label{ch:nanojets}
    
    In the simulations of the steady and stochastic heating, we assume an ideal MHD with a valid frozen-in condition, where the resistivity (magnetic dissipation) is negligible. However, due to the fact that we are solving continuous equations onto a discrete grid, the effects of numerical resistivity may appear and as a result, we can expect resistive processes to occur (more details on the influence of explicit physical resistivity $\eta$ in these simulations can be found in Sec.~\ref{ch:resisitiv_mhd}). 
    
\begin{figure}
        \centering
        \includegraphics[width=0.9\hsize]{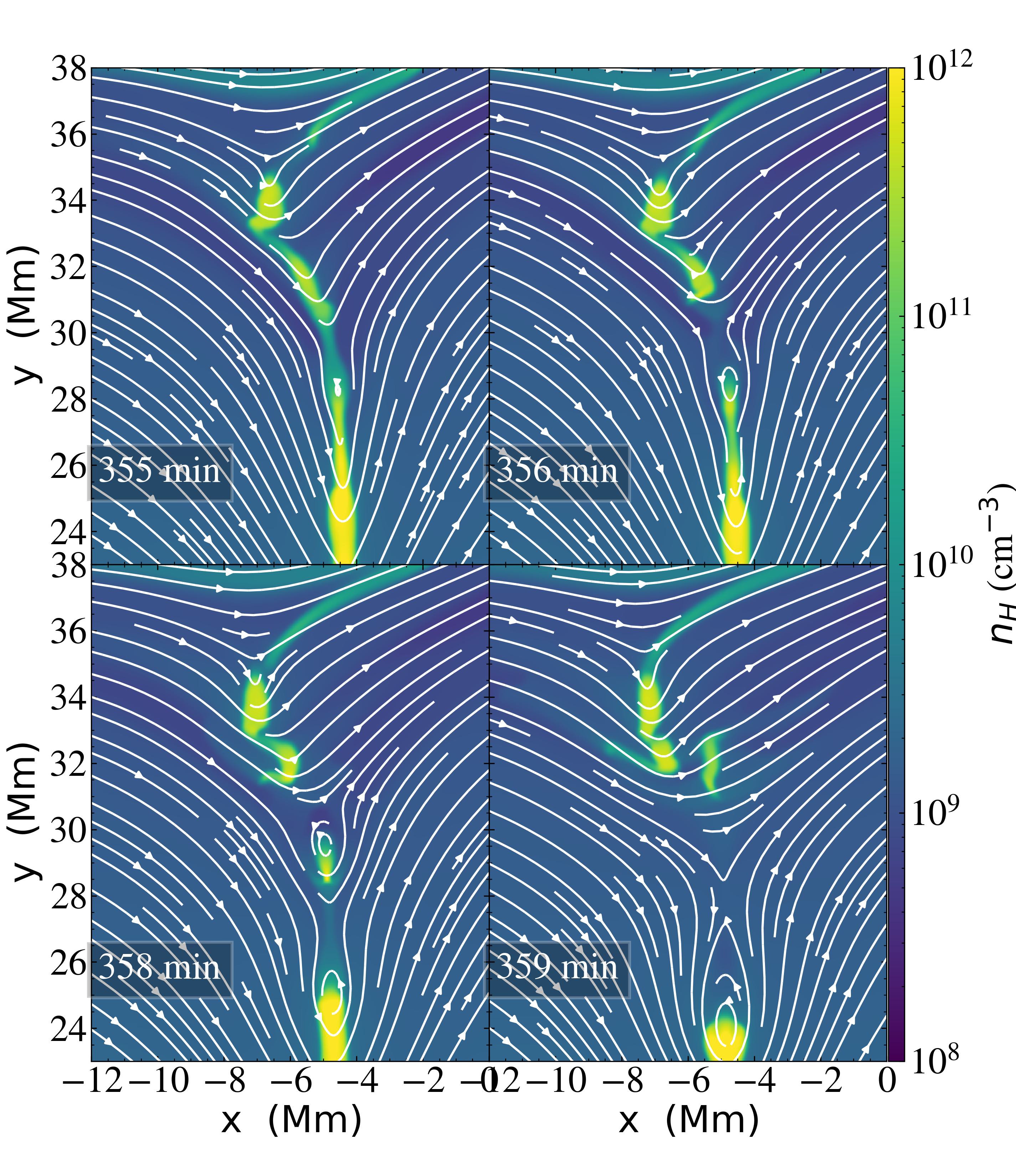}
        \caption{Distribution of $n_H$ in the region of the reconnection.}
        \label{fig:nanojet_multipanel}
    \end{figure}

    In the steady heating case, as shown in the snapshots of Fig.~\ref{fig:nanojet_multipanel} we observe first the main prominence body reconnecting at about 356\,min, separating away from a condensation on top of it. After that, another reconnection happens, where a smaller blob is seen on the panel at 358\,min. The steady heating drives enough evaporation to cause a high density prominence that, due to its large mass, inevitably dips the field lines and results in reconnection  \citep[a similar approach was used in][to create plasma droplets to explain hedgerow prominences]{Haerendel2011}. The particular 2.5 dimensionality of the system allows a relatively easy stretching of the poloidal magnetic field downwards and the consequent appearance of poloidal null points and reconnection. We observe a plasma blob shooting upwards as a result of this reconnection. The blob in our case is unidirectional and initially travels perpendicularly to the field lines, but soon after, it falls into a magnetic dip, follows the field lines and merges again with a plasma blob lingering on top of the main prominence body. After the reconnection, at 358\,min we noticed that the $v_y$ component suddenly increases up to $~\sim$70\,km\,s$^{-1}$. That makes the $v_{tot}$, which we measure to be $~\sim$80\,km\,s$^{-1}$, predominantly determined by the $v_y$ component. In the region between 26 and 28\,Mm of height, before reconnection happens, we had prominence material with temperatures of the order of 10$^4$\,K. After the reconnection, the temperature there increases to coronal values, greater than 1\,MK. Considering that the difference between consecutive snapshots shown here is around 85\,s, the jump in temperature we see does not happen instantly. Rather, we can expect sequential brightening in different AIA channels corresponding to the increase in temperature (from 304\,\AA, over 171, 193 to 211\,\AA). 
    Further on, the length of the ejected blob, looking at the temperature values of the order 10$^4$\,K, is approximately 1\,Mm, and the width is approximately half of the length. If we take the length and the width with $\rho\approx 6\times10^{-13}$\,g\,cm$^{-3}$ and $p_{th}\approx 1.5$\,dyn\,cm$^{-2}$ we can calculate kinetic ($0.5\rho v^2$) and thermal ($\frac{p}{\gamma-1}$) energies. The values are 1.5$\times$10$^{25}$\,erg and 1.76$\times$10$^{24}$\,erg respectively, which is in the range of nanoflare energies. To show in greater detail the exchange of energy happening in the area of the reconnection (for $y\in[23, 38]$\,Mm and $x\in[-2, -10]$\,Mm) we plot the change of magnetic, thermal and kinetic energies as weighted by the cell volume,
    \begin{equation}
        E_V = \frac{\int E dV}{\int dV}\,.
    \end{equation}
    \begin{figure}
        \centering
        \includegraphics[width=\hsize]{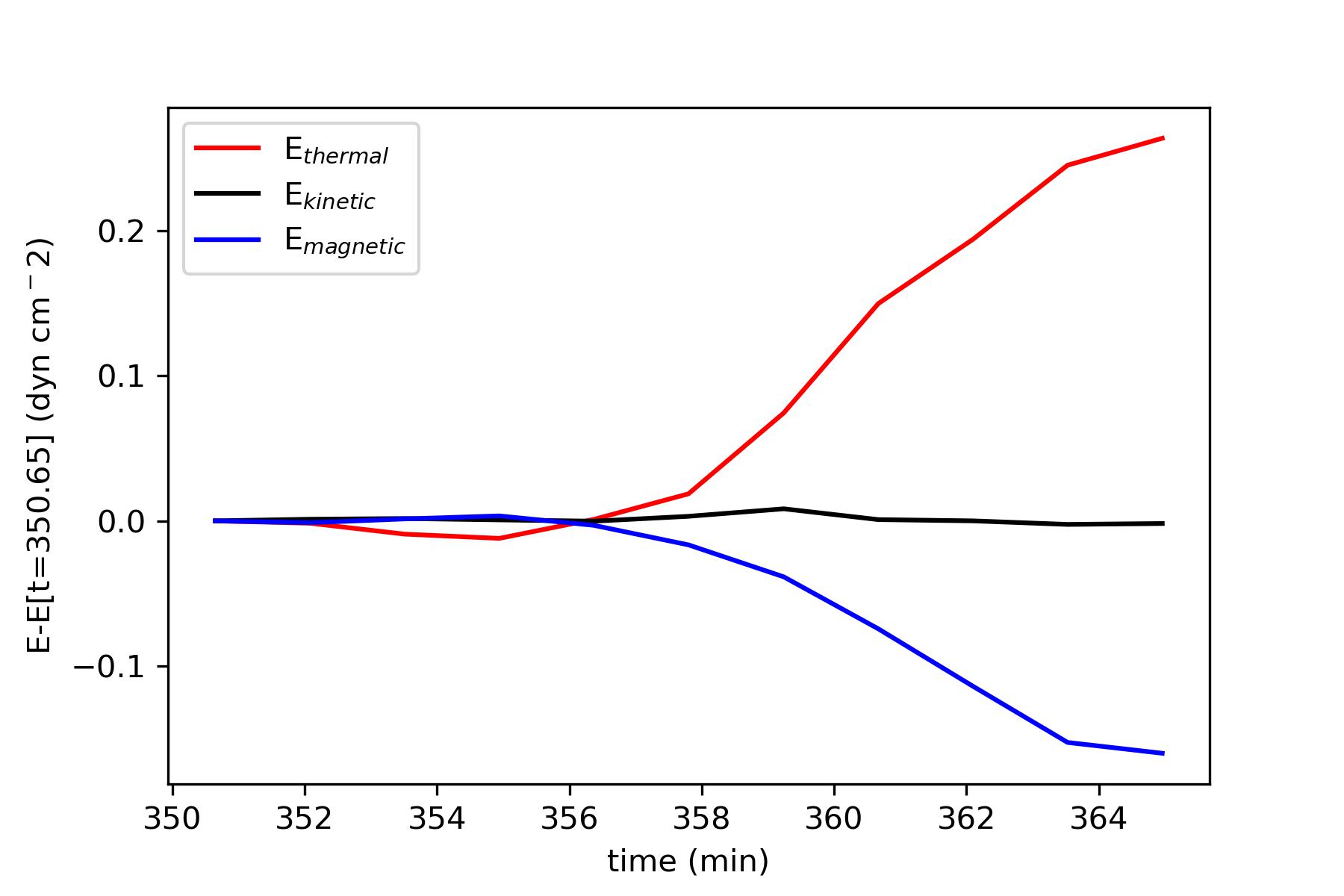}
        \caption{Thermal, kinetic and magnetic energy weighted with the volume and with respect to the energies at $t=350.65$\,min.}
        \label{fig:nanojet_energies}
    \end{figure}
    Figure~\ref{fig:nanojet_energies} shows a change in energy from a moment prior to reconnection ($\sim$350\,min). We see a clear exchange of energy from magnetic to thermal after the moment the reconnection happens ($\sim$356\,min). Since the reconnection is related to an increase of magnetic tension in the field lines, prior to $t=356$\,min we see a slight increase in magnetic energy, that drops as a result of field lines relaxing as they reconnect, followed by an increase in temperature and consequentially an increase in thermal energy.  
    
    Many characteristics of this event strongly resemble the nanojets observed by \cite{Antolin2021} and \cite{Sukarmadji2022}. In the case \cite{Antolin2021} reported on, the jet’s axis and trajectory are perpendicular to the magnetic field lines. In the follow-up work by \cite{Sukarmadji2022}, they define perpendicular ejection, fast velocities (100-200\,km\,s$^{-1}$) and the field line splitting after the nanojet, as useful guides for identifying such events. They describe a brightening in the location where the nanojet forms, followed by an ejection (nanojet) perpendicular to the field lines. Different observations they analysed do not always appear in the same AIA channels but rather, can appear in different channels for different observations, which is undoubtedly a consequence of their multithermal nature. They report on different cases in which the nanojet appears either individually or when there are multiple events observed, in what they call a cluster of nanojets. The length and the width of the plasma blob we measure in our simulation match very well to the average lengths and widths that were reported by \cite{Sukarmadji2022}, (1815$\pm$133)\,km for the length and (621$\pm$107)\,km for the width. Moreover, the kinetic and thermal energies we calculated are of the same order of magnitude as calculated by \cite{Antolin2021} and \cite{Sukarmadji2022}. Even though it exhibits all of the characteristics of nanojets, it is, by their definition, not a nanojet. Nanojets in \cite{Antolin2021} are defined as a result of small-angle reconnection. The type of reconnection creating the event in our simulation has clearly two field lines that are in the opposite direction, rather than at a small angle. Another possible explanation of the reconnection creating nanojets is the particular reconnection that was described by \cite{Kumar2023} for the same prominence as in \cite{Antolin2021}. They described the nanojets in the context of a failed filament eruption, where all the small jets are a consequence of the reconnection in the breakout and flare current sheets.

\subsection{Resistive MHD - How different $\eta$ affects the evolution}
\label{ch:resisitiv_mhd}

    Reconnection and slippage present important processes in the solar corona. The temperatures inside prominences are on the same order of magnitude as chromospheric temperatures, and the plasma is similarly partially ionised. Resistivity, hence, plays a role, and the interplay between processes such as reconnection versus plasma slippage \citep{Low2012a, Low2012b, Low2014} strongly affect the dynamics we observe. The plasma can either reconnect and in an energetic way cross field lines, or it can merely slip through, not causing any change in the field's topology \citep{Low2012a, Low2012b, Low2014}. To investigate how much the ideal MHD evolution of our two simulations differs in comparison to a resistive MHD, we explored the resistive aspects of the steady and stochastic heating prominences. We turned on resistivity (global and equal everywhere) at the 338th min in the steady heating case, just prior to the reconnection event already described in Section~\ref{ch:nanojets}. As for the stochastic heating, we did the same at $t=176$\,min. This particular moment was chosen, as by then a heavy blob was already seen forming (marked with an orange circle on the third panel of the bottom row in Fig.~\ref{fig:steady_stochastic_snapshots}), heavy enough to dip the field lines, and potentially cause a reconnection event. From those moments in the two simulations, we resumed the run for about 35\,min with resistivity turned on. We consider the following evolution for two different values of resistivity, $\eta_1=0.002$ and $\eta_2=0.0002$ \citep[dimensionless, with $\eta_{unit}=1.6282 \times 10 ^{-4}$\,s, same as in ][]{Zhao2022}.
    \begin{figure}
        \centering
        \includegraphics[width=\hsize]{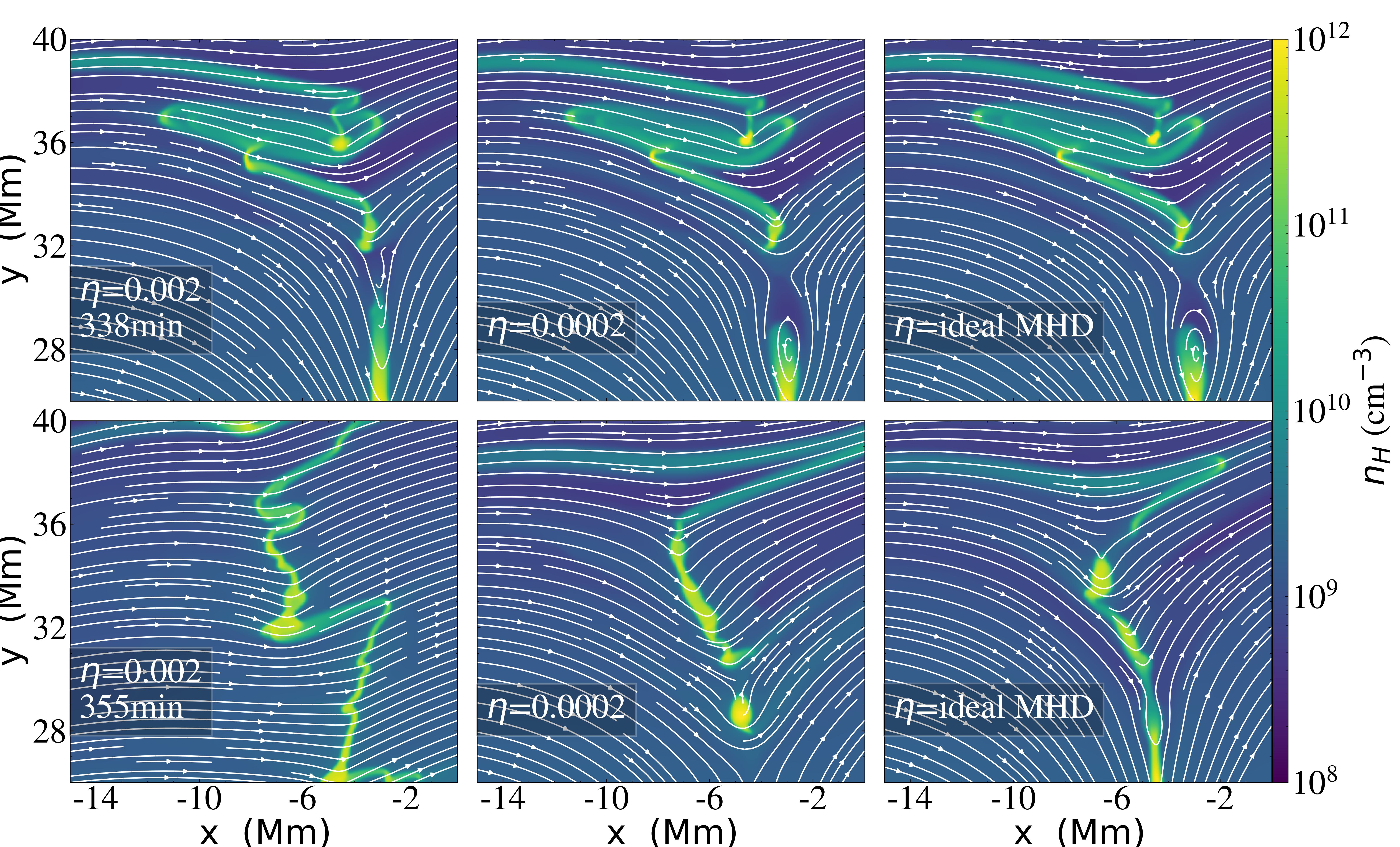}
        \caption{Distribution of $n_H$ for the case of steady heating with field lines over-plotted in white.}
        \label{fig:nH_etas_steady}
    \end{figure}
    \begin{figure}
        \centering
        \includegraphics[width=\hsize]{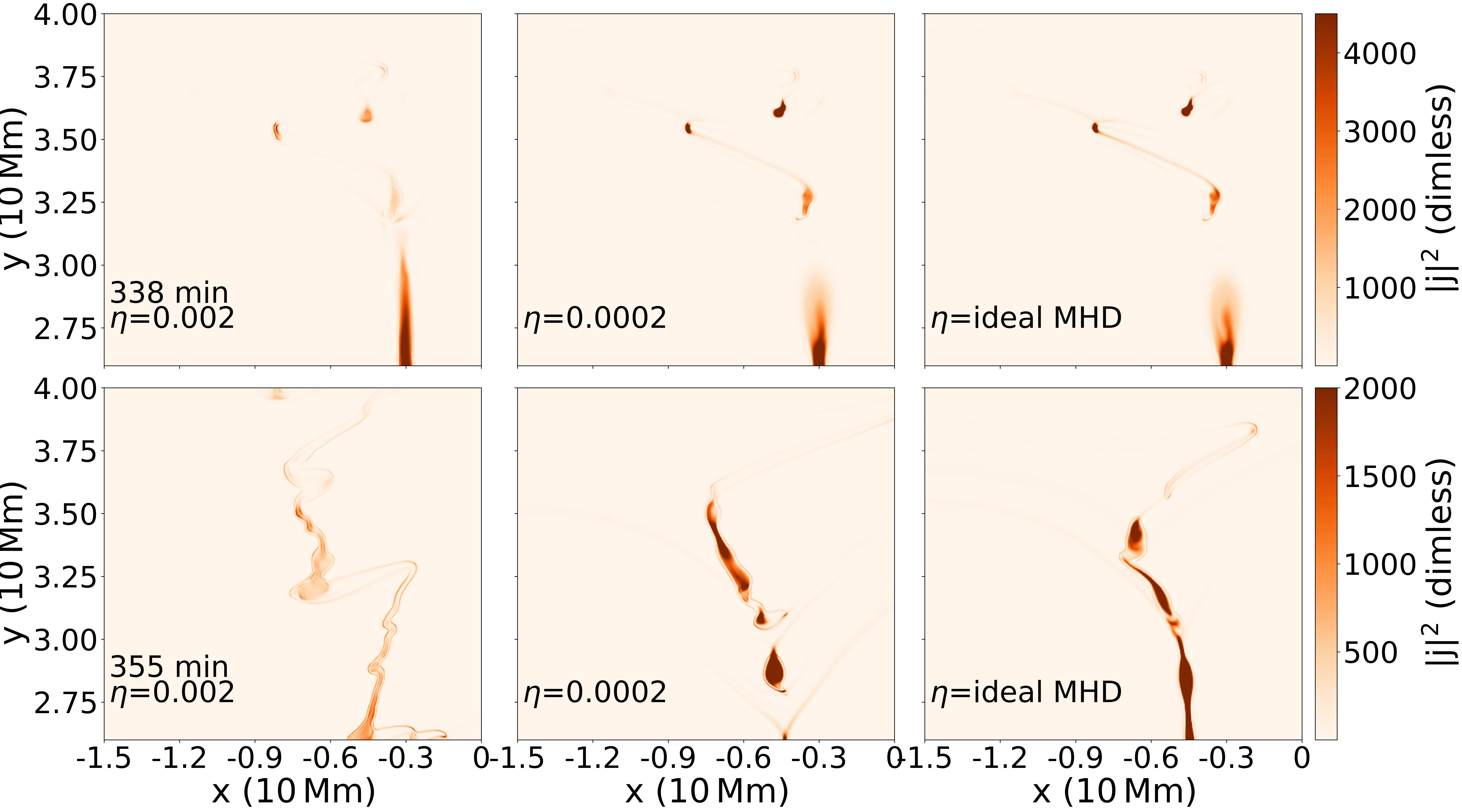}
        \caption{Distribution of $j^2$ for the case of steady heating (showing the same moments as in Fig.~\ref{fig:nH_etas_steady}).}
        \label{fig:current_steady}
    \end{figure}

    The top panels in Fig.~\ref{fig:nH_etas_steady} show the number density distribution at $t=338$\,min, a snapshot about 86\,s after we turn the resistivity on in the steady heating case. The bottom panels are taken 17\,min later, showing how much in the elapsed time the evolution already differs between $\eta_1 = 0.0002$, $\eta_2 = 0.002$ and the ideal MHD. Figure~\ref{fig:current_steady} shows the same but with the distribution of $j^2$. Initially, as seen in the top panels, there is no significant difference between the ideal MHD and $\eta_1 = 0.0002$, and both show in certain locations greater current values in comparison to $\eta_2 = 0.002$ (leftmost panel). Because the resistivity is the highest in this case, it allows for more efficient dissipation of magnetic energy as joule heating is proportionally amplified. In other words, more magnetic energy is converted into thermal, unlike when resistivity is small or non-existent. 17\,min after we turned on the resistivity the situation has significantly changed. The bottom panels of Fig.~\ref{fig:nH_etas_steady} show how the two runs with resistivity enabled, evolved differently in comparison to the ideal MHD. The connection between different blobs differs, as well as the shape of those condensations. The condensations in the $\eta_2 = 0.002$ are all still connected into one thread, as the enhanced diffusion inhibits the localisation of large currents. $\eta_1 = 0.0002$ shows already disconnected blobs, and we know the same also happens in the ideal MHD case. Smaller resistivity (or in the case of an ideal MHD a non-existent one) allows for a build-up of magnetic tension, meaning the reconnection happens later but also the velocities observed after such reconnection are greater. Differences in the value of $j^2$ are now a lot more obvious between $\eta_2 = 0.002$, $\eta_1 = 0.0002$ and the ideal MHD as clearly seen in the bottom panels of Fig.~\ref{fig:current_steady}. In the full $\sim$35\,min that we run the simulations with resistivity, the low value of $\eta_1 = 0.0002$ did not significantly change the evolution in comparison to ideal MHD. On the other hand, in the $\eta_1 = 0.002$ case, the main prominence body completely slipped through the field lines and drained into the chromosphere, hence differing in its evolution significantly in comparison to the ideal MHD. 

    We did the same analysis with the stochastic type heating. However, in that case, there are no obvious reconnection events as in the steady heating case. At $t = 175$\,min we noticed more mass piling in a particular blob-like region, creating its own dip in the field line. We marked that as a potentially good moment to resume the run with resistivity on. Similar to the steady heating case, we see that the higher the resistivity, the more current is dissipated and turned into ohmic heating. Regardless, significant topological differences between an ideal MHD and a resistive one are not as obvious as they are in the steady heating case. Whilst the presence of resistivity might change the energy balance by heating the plasma and causing a stronger loss of energy via optically thin radiation, we can say that in the $\sim$35\,min the resistivity was present, it did not influence the global evolution significantly. Strong flows along the field lines in the stochastic heating case are constantly forcing the plasma to move and do not allow for any noticeable reconnection and/or slippage. 
  
    Given that the prominence plasma in reality is not fully ionised and it is cold enough that neutrals are expected, resistivity is expected. However, since we do not know the exact values of $\eta$ (and it most likely differs for different regions of the same prominence, i.e. filament) it is also difficult to predict how the plasma behaves; if we can expect the prominence mass to preferentially cause reconnection events, simply heat, or resistively slip, is still not known. What is more, if we compare the influence of higher resistivity in the steady heating case versus the stochastic one, we see that resistivity and its effects are less important when there is a presence of strong flows. In stochastic heating, the flows of evaporated plasma from the footpoints do not allow for plasma to accumulate at any location long enough to form strong cross-field gradients, in which case the strong resistivity has no effect, and we do not see any slippage. In conclusion, both resistivity and flows play a role in the behaviour of plasma on a large scale, be it predominantly horizontal or vertical motion. What is more, both also affect the behaviour of plasma on a small scale, the likelihood of reconnection happening and consequently, the chance that we see nanojets (appearing individually, in a cluster or appearing at all).

\subsection{Synthetic spectra - What can we anticipate from observations?}

    To further extend our work in the direction of observational aspects, we follow the recent work of \cite{Jenkins2023}. We used their methods to recreate synthetic spectra of \halpha, \Caii{~H\&K}, \Caii{~8542} and \Mgii{~h\&k} using the parameters of our simulation, height, temperature, density, and line-of-sight (LOS) velocity. Using their methods means we used \textit{Lightweaver}, where \textit{Lightweaver} adopts a 1.5D assumption thus far, where one can account for centre-to-limb variations, but all NLTE radiative transfer aspects are dealt with in height only. We were particularly intrigued by the reconnection described in Sec.~\ref{ch:nanojets}. 

    In an extension to the methods implemented in \citet{Jenkins2023}, we simultaneously iterate the electron number density, at each iteration step, from an initial NLTE distribution that is commonly assumed fixed \cite{Heinzel2015a}; the \Caii~and \Mgii~resonance lines are known to be strongly influenced by the electron number density. The necessary Newton-Raphson iteration procedure is already incorporated into \textit{Lightweaver} (\texttt{ConserveCharge = True} in the \texttt{Context} constructor), where here we also maintain the initial pressure conditions by subsequently modifying the total Hydrogen number densities and then all other populations by the equivalent ratio. This approach was originally omitted from the work of \citet{Jenkins2023} as the increase in computational cost proved prohibitive for the number of atmospheres considered there. In general, we find the stratification of $n_e$ after iteration to be of larger magnitude, with a relative difference of order +10\% to +50\% in the coronal and prominence portions, respectively. The largest differences were found in the deepest regions of the prominence containing very large total hydrogen number densities of order 10$^{11}$\,--\,10$^{12}$\,cm$^{-3}$.
    \begin{figure*}
        \centering
        \includegraphics[width=\textwidth]{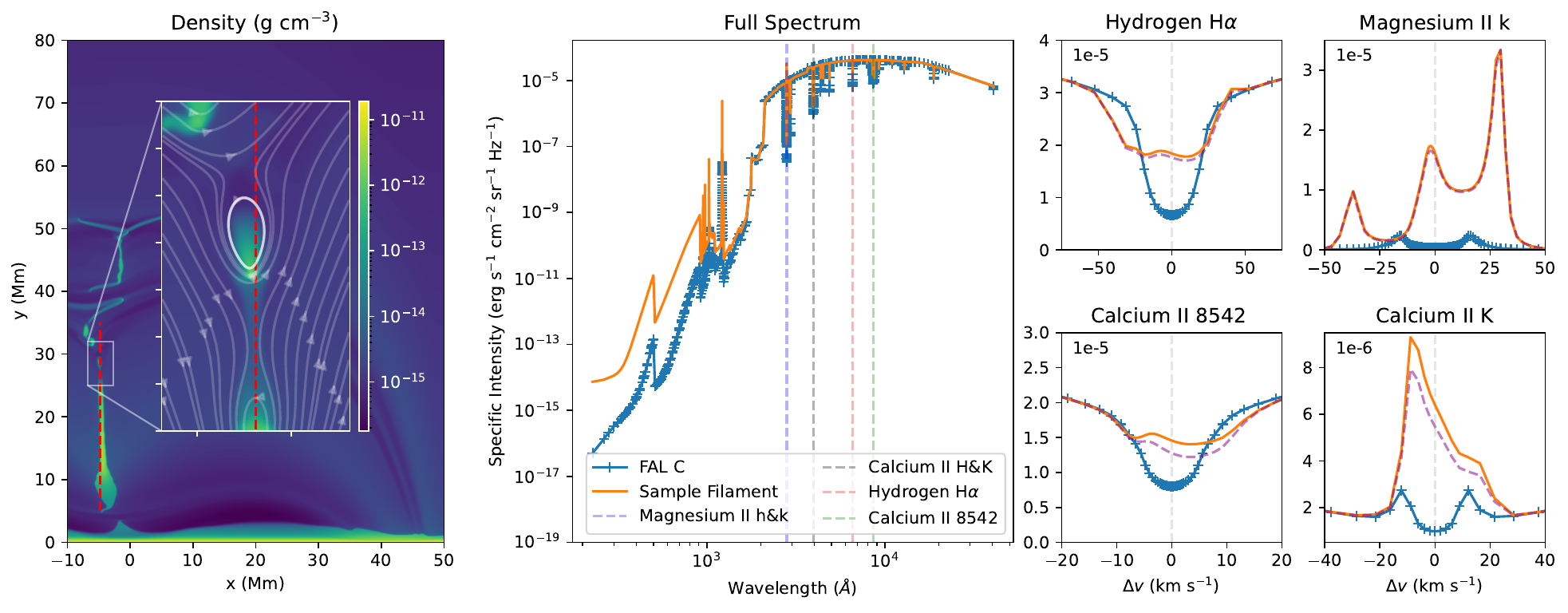}
        \caption{On the left is the density in the domain at the moment of reconnection ($t=357.64$\,min) for which we also plotted the right panels. The panel in the middle shows the full spectrum together with the FALC for the full range of wavelengths we consider within \textit{Lightweaver}. The four panels on the right zoom in on the spectral lines, \halpha, \Caii{~8542} and \Caii{~K} and \Mgii{~k} lines. The profiles constructed assuming iterated and fixed $n_e$ are shown in solid-orange and dashed-purple, respectively.}
        \label{fig:spectral_lines}
    \end{figure*}

    Figure~\ref{fig:spectral_lines} shows a comparison of the full spectrum as created by \textit{Lightweaver} from the parameters of our steady heating simulation (solid-orange versus dashed-purple showing when $n_e$ iteration is not taken into account) together with the common empirical chromospheric model, FAL-C \citep[blue,][emplyed here as the chromospheric illumination]{Fontenla1993}. The panel on the left shows the moment when and where we observe the spectra. The spectra are created from a filament point of view, that is the LOS is top-down, along the red dashed line seen in the left panel of Fig.~\ref{fig:spectral_lines}. The snapshot was taken at the moment when the plasma blob was already reconnected at $t=357.64$\,min. In the middle panel, displaying the full spectrum, the EUV region is particularly interesting as it shows a jump in comparison to the FAL-C model. In other words, the EUV continuum increases are indicative of a reconnection event that could be seen in SDO/AIA. In the four panels on the right we see \halpha, \Caii{~8542}, \Caii{~K} and \Mgii{~k} lines and how they changed due to the reconnection in comparison to the standard FAL-C model. The \halpha{} linecore is enhanced and contains a second, strongly blue-shifted component, while \Caii{~8542} does not respond so clearly to the velocity change it does also show a stronger emission from the prominence in comparison to the chromospheric profile. The \Mgii{~k} and \Caii{~K} both show a similar response as they both exhibit signs of strong emission where there would be absorption under quiescent conditions such as the FAL-C. In order to explain these changes the contribution function is a helpful tool as it tells us how much a local voxel along the LOS contributes to the intensity we observe. The contribution function is calculated according to \cite{Carlsson1997} \citep[also exactly the same as][]{Jenkins2023},
    \begin{equation}
        C(\nu,z_t) \equiv \frac{dI}{dz}\bigg|_{z=z'} = \frac{\chi_{\nu}^{tot}(z')}{\tau_{\nu}(z')}S_{\nu}(z')\tau_{\nu}(z')e^{-\tau_{\nu}(z')}\,,
    \end{equation}
    where $z_t$ represents the top of the red dashed line in Fig.~\ref{fig:spectral_lines} (left panel). 
    \begin{figure*}
        \centering
        \includegraphics[width=\textwidth]{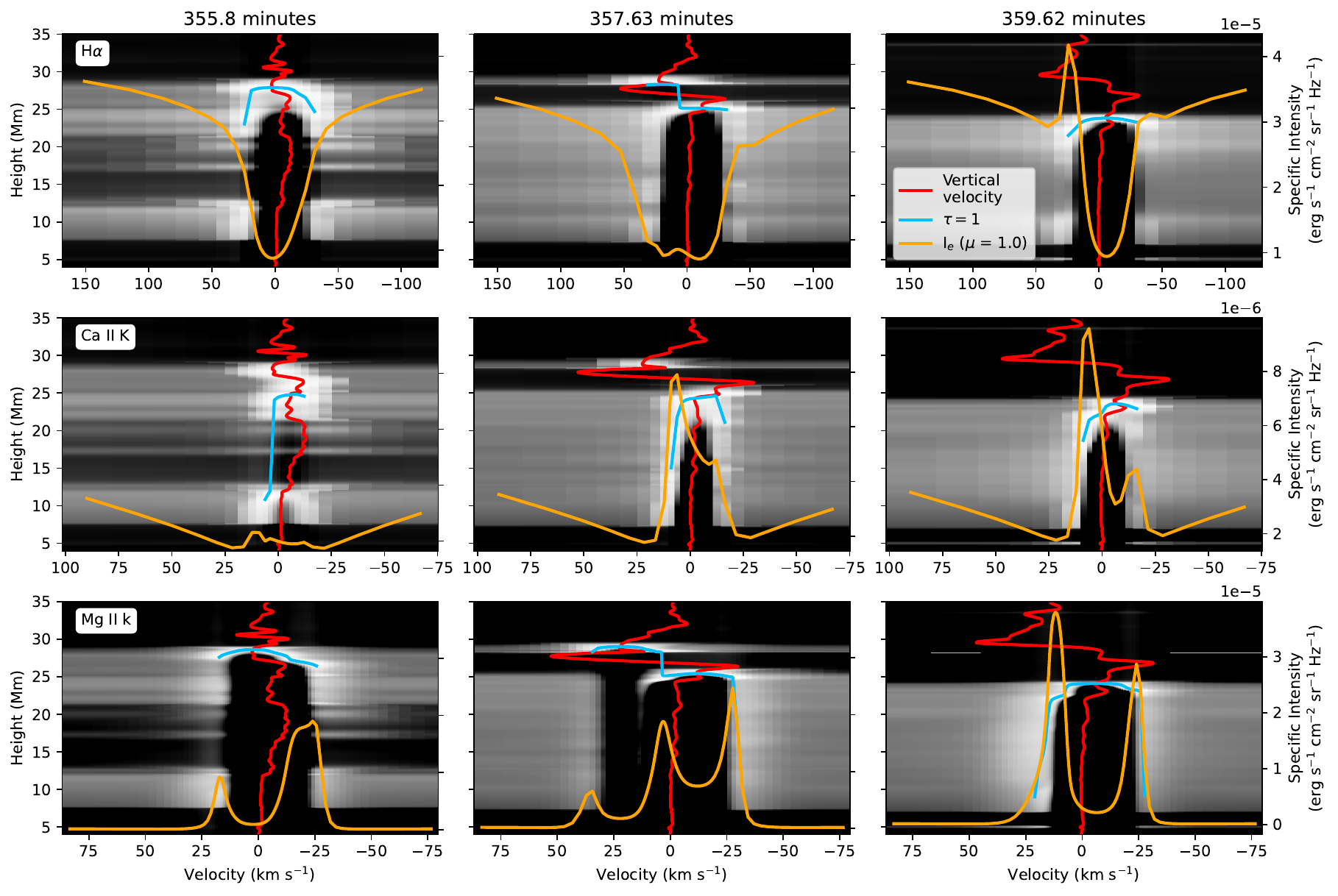}
        \caption{Contribution functions of H$\alpha$, and \Caii{~K} and \Mgii{~k} lines in three different moments (before, during, and after the reconnection event), together with overplotted profiles of the $v_y$ component (red), position for the $\tau=1$ layer (light blue), and the specific intensity for each spectral line (orange).}
        \label{fig:contribution_functions}        
    \end{figure*}
    We plot the contribution function for \halpha, \Caii{~K} and \Mgii{~k} lines in Fig.~\ref{fig:contribution_functions} at three different moments, which are before (left column), just after the reconnection (middle column) and still after the reconnection when the reconnected plasma blob is outside the LOS (right column). Given the large model densities, the main absorption we see in \halpha~comes from the prominence-corona-transition-region (PCTR). Before the reconnection happens (left column), it originates from the PCTR of the main prominence body. Already straight after the reconnection the main contribution to \halpha~comes from the PCTR of the ejected blob. As the $\tau$ = 1 line is narrowly concentrated to the height where we see the plasma blob we can easily relate it. That then also explains why the \halpha~line core is strongly blue-shifted, as it is responding to the upwards velocity of the blob. In the middle column, past the line core on the right side, there is an additional contribution that makes the line core of \halpha~narrower. That is, the $\tau = 1$ layer is now located through multiple layers in the range of $\sim$25 to 29\,Mm of height. However, we see that the main contribution comes from the bottom of the blob and the top of the main prominence body; together their PCTRs cause the slight line core enhancement. There is also an additional feature of the \halpha~spectra. It is the peak clearly seen to the left of the line core in the final timestep. We speculate that it results from a contribution of other lines; the UV Mg triplet lines, that can be found between Mg h and k and in most cases in absorption, are found here to be in emission \citep{Pereira2015}. Hence, we propose that those lines are contributing to the peaks in the \halpha~line wings (Pereira, M. D. T. private communication).

    Moving on to \Caii{~K} and \Mgii{~k} lines. Unlike \Mgii{~k} and \halpha, \Caii{~K} $\tau=1$ layer is in a wider region of significant contribution. In all three columns, we see the $\tau=1$ layer at the height of $\sim$24\,Mm (what becomes the prominence PCTR after the plasma blob reconnects). In that region, as the reconnection evolves, there is a sequential change in temperature. As a result, \Caii{~K} line is formed inside those layers that then represent emission rather than absorption, which would, in normal circumstances result from deep down inside the prominence. The velocity gradient also partially contributes to the emission, though most of the contribution comes from the temperatures found in the prominence PCTR. Although the contribution function of \Caii{~K} also peaks within the PCTR of the reconnected blob, it appears the optical depth is not high enough here for the spectral line to record this.

    Lastly, the \Mgii{~k} line, similarly to \halpha, is responding to the double PCTR that appears, one at the bottom of the blob and the other at the top of the main prominence body. This results in the fact that initially (left column) we see the peaks growing in intensity and moving away from the centre. As the blob moves upwards and the main prominence body is pushed down, the left peak in \Mgii{~k} becomes strongly blue-shifted and the right peak becomes red-shifted. When the two PCTRs are fully formed (middle column) a third peak appears between the two \Mgii{~h} peaks. It represents emission, resulting from the temperatures found at the aforementioned PCTRs. In the third column, the blob is already out of the LOS we observe and hence we only see the usual \Mgii{~h} peaks as formed inside the prominence and its PCTR that continue to have large intensities (in comparison to the FAL-C as seen on Fig.~\ref{fig:spectral_lines}).
    \begin{figure}
        \centering
        \includegraphics[width=\hsize]{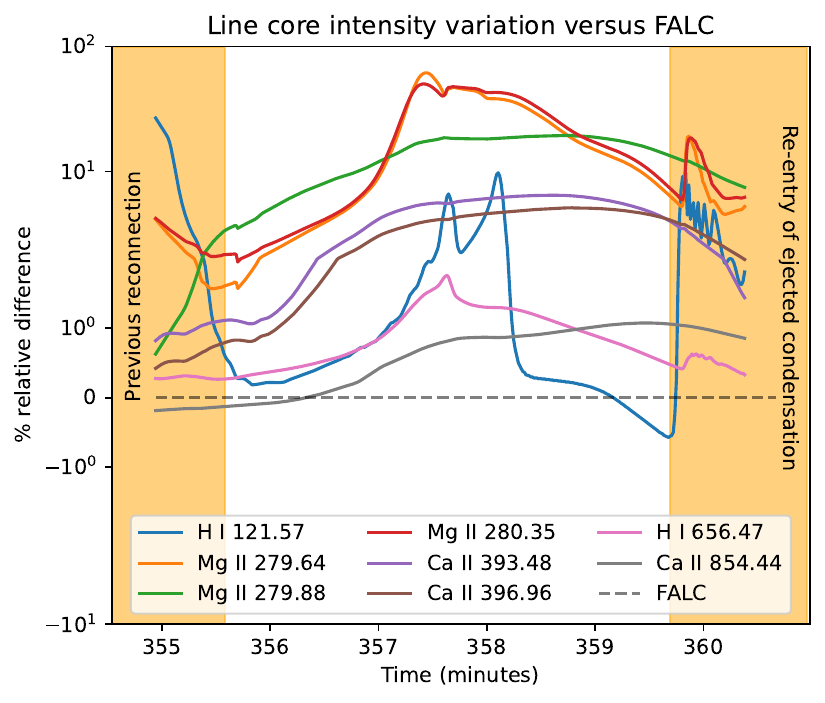}
        \caption{Relative difference (in percentage) of the line cores of various elements versus the FAL-C. The orange-shaded regions mark complex events seen before and after reconnection.}
        \label{fig:line_core_intensity_variation_vs_FALC}
    \end{figure}
    
    In order to compare the line core intensities among each other and with FAL-C, and additionally how they change during the whole time period of reconnection, we refer to Fig.~\ref{fig:line_core_intensity_variation_vs_FALC}. In this way, we can follow the change in time clearly and also see in more detail how the reconnection causes each line core to deviate from the standard FAL-C model. The evolution here is non-linear and the situation is complex. First, there is another reconnection happening before the one we study, that detaches the main prominence body from the structure on top. Second, the blob does not follow a straight path after reconnection, but it exits and re-enters our LOS. All those events leave complex signatures in the spectral lines and they are labelled in Fig.~\ref{fig:line_core_intensity_variation_vs_FALC} by an orange shaded area. We do not go into detail describing those events but rather focus on the changes we see in the lines around the time the reconnection happens ($\sim$356\,min to $\sim$359\,min). The \Mgii{~h\&k} (279.64 and 280.35\,nm, red and orange) clearly show a much higher intensity than the FAL-C. The reason is that the LOS passes through the PCTRs that contain temperatures that strongly contribute to the \Mgii~emission. That is additionally supported by the UV Mg triplet line (\Mgii{~279.88}\,nm, green line) going also into emission during the whole period the reconnection is happening. Similar behaviour to \Mgii~lines is exhibited by \Caii{~H\&K} lines (393.48 and 396.96\,nm, purple and brown). They also show strong emissions in comparison to FAL-C. Moreover, as the \halpha~line (pink line) is strongly blue-shifted by the upwards motion of the plasma blob, looking at the line core, it also exhibits higher intensity than the FAL-C. The \Caii{~IR~854.2}\,nm line (854.44: grey line) follows similar behaviour to the resonance counterparts, albeit with a weaker enhancement owed to its smaller opacity and hence deeper formation height that is far from the reconnection site. The last line plotted in Fig.~\ref{fig:line_core_intensity_variation_vs_FALC} is H~I 121.57\,nm (blue line) which is the closest to the EUV part of the spectrum, and by far the most optically thick line that we consider here. It nonetheless appears to respond most dramatically to the reconnection event, showing an impulsive increase in intensity. This can be attributed to its formation in the outermost, and hence the hottest, portions of the PCTRs found along the LOS. 

%-----------------------------------------------------------------

\section{Summary and conclusion}
\label{ch:conclusions}
    In this work, we described the properties of two prominences formed by a steady and a stochastic type of localised heating. Both types of heating were previously considered separately in other studies \citep[][respectively]{Keppens2014, Jercic2023}, but here are revisited within a single, high-resolution model and directly compared therein. The numerical details of the two runs we did here were the same, nonetheless, we see the global characteristics of the two prominences differing significantly. We summarise here the main differences that can be directly related to the different types of localised heating and in turn to observational features.
   \begin{enumerate}
      \item In the case of steady heating, the prominence that forms is compressed by the constant flows driven by the particular steady heating. As more plasma is constantly condensing onto the prominence and due to the flows limiting the prominence expansion in width, the formed prominence is a very dense and massive one (more than 2$\times10^{5}$\,g\,cm$^{-1}$). Due to the large mass that accumulated, the prominence heavily weighs on the field lines, causing the field lines to dip and eventually reconnect, creating tiny flux ropes. Additionally, as the prominence is strongly constrained by the steady heating it shows limited dynamics. The prominence itself is predominantly static and does not exhibit significant motion on a large scale. Also present is coronal rain. It forms on top of the main prominence body and is strongly influenced by the ram pressure of the flows coming from the footpoints resulting in the thin shell instability. The plasma, then fragmented, rains down along the field lines.
      \item On the other hand, the stochastic type of heating results in a prominence with prevailing thread-like structures. The condensations form randomly all over the domain and are influenced by the flows coming from the specific stochastic heating at the footpoints. As the condensation forms it gets pushed along the field line and on many occasions, it quickly drains back into the chromosphere (coronal rain). Besides drainage, different threads also coalesce and merge and persist in the domain for a long time (longer than an hour). As such there is a localised build-up of condensed mass. Such build-up of mass that we see in our simulation also eventually drains. As the heating is constantly present we expect similar processes to repeat and more dense blobs to form later on. Unlike the steady heating case, the prominence here is highly dynamic, reflecting directly the properties of stochastic heating.
      \item Comparing our stochastic heating simulation with observations allowed us to relate the transverse oscillations we see in the domain with the particular motion of the threads and how they affect the field lines, which are furthermore affected by the flows coming from the localised heating. As such the oscillations we measure are directly influenced by the threads and indirectly by the stochastic heating at the footpoints.
      \item The striking difference between the resulting dynamics of the two prominences, governed by their particular heating, affects also their potential to exhibit reconnection. As the heavy, steady prominence weighs the field lines, it brings them to a favourable position for reconnection. On the other hand, the stochastic heating threads are so dynamic that they do not linger long enough at one location to affect the field lines in a similar manner. 
      \item Including resistivity in the MHD equations results in significant differences in the evolution of the steady heating case. With uniform resistivity, additional diffusion is now present, allowing the plasma to simply slip across the field lines. In contrast, in the case of stochastic heating, not even relatively high resistivity causes any noticeable topological changes in comparison to the ideal MHD (during the limited time we introduce resistivity in the simulation).
      \item Observing the filament spectra provides us with various details on the evolution that we can directly relate to the changes in the parameters that we measure within the simulation (such as temperature and velocity as the most significant contributors). Regardless, the full potential of creating synthetic spectra from the simulation is not be fully explored and achieved until more detailed comparisons with real observations are done. A recent study by \cite{Panos2023}, done using machine learning on spectra observed by IRIS, showed that the most significant precursors of flares are high triplet emission and core intensity, irregularly shaped profiles, broadening of the spectral cores, single peaks in the spectra as well as the flows seen as extended red and blue wing emission. From Fig.~\ref{fig:spectral_lines} and~\ref{fig:contribution_functions} we can see similar characteristics appearing in our spectra which we can clearly relate to the reconnection that was happening at that moment of the simulation. Nevertheless, more work is indeed needed in the 1.5D synthesis in order to understand the full imprint of the prominence on the spectra and to make proper comparisons of simulations and observations. 
   \end{enumerate}
   The 2.5D type of simulations make it convenient to study different processes as they come as close as possible to a 3D simulation while still allowing extremely high resolutions for very large domains. Despite their advantages, there are still some disadvantages that one needs to be wary of. For example, our reconnection, even though a result of the heavy prominence mass, is also influenced by the fact that our $x$ and $y$ magnetic field components are quickly decreasing with height. As a result, they cannot offer much support to the heavy plasma. On the other hand, the $z$ component of the magnetic field, which is constant with height and strong enough, is invariant in the direction into the plane of the simulation. Hence, it cannot offer the full support as in a full 3D simulation where the dips may form in the dominant $z$-component; and this affects how easily a reconnection might happen. In the future, we plan to extend this work by doing simulations in 3D and further explore this and similar setups. In that way, not only are we able to consider the changes in the now invariant component, but also allow for a different view of the reconnection we touched upon here. A view from different angles will prove important for the synthetic spectra analysis and comparison with observations of prominences and filaments in the solar corona. 

\begin{acknowledgements}
      VJ acknowledges funding from Internal Funds KU Leuven and Research Foundation – Flanders FWO under project number 1161322N. RK and JMJ are supported by the ERC Advanced Grant PROMINENT and an FWO grant G0B4521N. This project has received funding from the European Research Council (ERC) under the European Union's Horizon 2020 research and innovation programme (grant agreement No. 833251 PROMINENT ERC-ADG 2018). This research is further supported by Internal funds KU Leuven, project C14/19/089 TRACESpace. Visualisations used \href{https://www.paraview.org/}{ParaView}, \href{https://www.python.org/}{Python}, \href{https://yt-project.org/}{yt} and \href{https://github.com/Richardjmorton/auto_nuwt_public}{Auto-NUWT}. The resources and services used in this work were provided by VSC (Flemish Supercomputer Center), funded by the Research Foundation - Flanders (FWO) and the Flemish Government. VJ wants to thank Dr Daye Lim for introducing her to Auto-NUWT.
\end{acknowledgements}

\bibliography{reference}{}
\bibliographystyle{aa}

\end{document}